\documentclass[mathleft
]{an}
\usepackage{graphicx}
\usepackage{times}
\usepackage{booktabs}
\usepackage{amsmath}
\usepackage{amssymb}
\usepackage{threeparttable}
\usepackage{hyperref}
\usepackage{natbib}
\bibpunct{(}{)}{;}{a}{}{,}

\begin{document}

\Pagespan{789}{}
\Yearpublication{2006}%
\Yearsubmission{2005}%
\Month{11}%
\Volume{999}%
\Issue{88}%

\title{Determination of a temporally and spatially resolved Supernova rate from OB-stars within 5\,kpc}

\author{J.G. Schmidt\fnmsep\thanks{Corresponding author:
  \email{Schmidt.Janos@uni-jena.de}\newline}
\and  M.M. Hohle
\and R. Neuh\"auser
}
\titlerunning{The SN rate in the Solar Vicinity}
\authorrunning{J.G. Schmidt, M.M. Hohle, \& R. Neuh\"auser}
\institute{
Astrophysikalisches Institut und Universit\"ats-Sternwarte Jena, Schillerg\"asschen 2-3, 
D-07745 Jena, Germany
}

\received{30 May 2005}
\accepted{11 Nov 2005}
\publonline{later}


\abstract{%
We spatially and temporally resolve the future Supernova (SN) rate in the Solar vicinity and the whole Galaxy by comparing observational parameters of massive stars with theoretical models for estimating age and mass and, hence, the remaining life-time until the SN explosion. 
Our SN rate derived in time and space for the future (few Myr) should be the same as in the last few Myr by assuming a constant rate.\newline
From BVRIJHK photometry, parallax, spectral type, and luminosity class we compile a Hertzsprung-Russell diagram (H-R\,D) for 25\,027 massive stars and derive extinction, and luminosity, then mass, age, and remaining life-time from evolutionary models.\newline
Within 600\,pc our sample of SN progenitors and, hence, SN prediction, is complete, and all future SN events of our sample stars take place in 8\,\% of the area of the sky, whereas 90\,\% of the events take place in 7\,\% of the area of the sky. The current SN rate within 600\,pc is increased by a factor of 5-6 compared with the Galactic rate. 
For a distance of 5\,kpc our sample is incomplete, nevertheless 90\,\% of those SN events take place in only 12\,\% of the area of the projected sky. If the SN rate in the near future is the same as the recent past, there should be unknown young neutron stars concentrated in those areas. 
Our distribution can be used as input for constraints of gravitational waves detection and for neutron star searches.
}
\maketitle

\section{Introduction}
The search for gravitational wave (GW) events and pulsars using \textit{Einstein@home}\footnote{see http://einstein.phys.uwm.edu} can work more efficient with spatial limits. 
The spatially and temporally resolved future Supernova (SN) rate is assumed to be roughly constant over some Myrs, and hence can indicate promising regions for the detection of probable GW sources such as precessing neutron stars (NSs) or binary NSs.
The probability to find young NSs, which might be precessing and radiating GWs due to ellipticity, is larger in areas with a local increased SN rate. These sources can be detected with GW detectors (such as adLIGO). All sky surveys like \citet{Aasi2013} could be reduced to smaller areas in the sky by using our results. Thus the sensitivity of GW searches could be increased significantly.\\
100 to 140 NSs younger than 4\,Myr should be expected within 600\,pc from population synthesis \citep{Popov2005}, but only 15 of them are known yet in the ATNF pulsar catalogue \citep{Manchester2005}.
The Galactic core collapse SN (ccSN) rate is estimated to $2.5^{+0.8}_{-0.4}$ events per century by counting SNe from Milky Way like galaxies \citep{Tammann1994}.
The amount of ${}^{26}$Al \citep{Diehl2006} implies a rate of (1.9$\pm$1.1) ccSNe per century in our Galaxy. 
\citet{Grenier2004} found 20 to 27 SNe per Myr for the Gould Belt \citep{Gould1879}, a local star forming region within 600\,pc \citep[for an historical overview see][]{Stothers1974}, by studying observed star counts. This local current rate is 5 to 6 times higher than the Galactic average.
A spatially and temporally resolved SN rate of 21$\pm$5\,SNe per Myr within 600\,pc was published by \citet*{Hohle2010}. They analysed the SN rate on the base of massive stars observed with both HIPPARCOS and 2MASS. 
\citet{Hohle2010} derive mass and age for SN progenitors by using different evolutionary models, and determine the remaining life-time.
Their sample is complete for stars brighter than the limiting magnitude of $V$=$8.0$\,mag \citep{vanLeeuwen2007} for HIPPARCOS. Thus all B4V and earlier stars within a distance of 600\,pc are included, assuming a mean interstellar extinction of $a_V$=1\,mag/kpc for the solar vicinity \citep{Lynga1982}.\\
In this work we enhance the analysis of the SN rate for more distant stars. We use additional and more recent databases to find more ccSN progenitors in the solar vicinity up to 5\,kpc.
Furthermore, while all spectral types in \citet{Hohle2010} are taken from the HIPPARCOS catalogue \citep{Perryman1997} or SIMBAD, we also use more recent spectral classifications. Each step of the analysis is reconsidered and improved.
Wolf-Rayet stars (WR) were not included in \citet{Hohle2010} but will be included in this work.
For a better evaluation of the numbers we calculate the completeness of our sample to specify the areas in the sky which should be explored in more detail in the future.

\section{The sample}
\label{sec:sample}
The base of the whole analysis is a set of potential SN progenitors ($M>8\,M_{\odot}$) that is queried for observational parameters.  
We select all ccSN progenitors regarding the spectral type classification (SpT): we cover all SpT for luminosity classes (LCs) I and II, all stars earlier than or equal to SpT B9 for LC III, and all known stars earlier than or equal to spectral type B4 with LC IV, V, or an unknown LC will be also included (\autoref{tab:selcriteria}).
\begin{table}
\centering
\caption{Selection criteria for the luminosity classes and the spectral types.}
\begin{tabular}{lccc}\toprule
luminosity class & I, II & III & IV, V, unknown\\ \midrule
spectral type    & all   & earlier/eq. B9 & earlier/eq. B4\\
\bottomrule
\end{tabular}
\label{tab:selcriteria}
\end{table}
The basic set of stars comes from a \textit{select by criteria} query in SIMBAD (see \autoref{sel:SIMBAD}), consisting of 18\,146 objects (as of 2013-03-04).
We expand the sample from SIMBAD with the catalogues in \autoref{Tab-A:Katalogerkennungen} using the VizieR Online Database \citep{Ochsenbein2000}.
There are 13\,741 additional stars fulfilling our selection criteria from \autoref{tab:selcriteria} plus 226 WR stars from \citet{vanderHucht2001}. Thus our sample expands to 32\,113 stars. Some catalogues are including bright stars not belonging to our Galaxy. We avoid an extragalactic overpopulation in our sample by excluding 7\,165 stars in the sky area of the Large and the Small Magellanic Cloud.

\begin{figure}
\includegraphics[width=\columnwidth, trim=8 8 7 17, clip=yes]{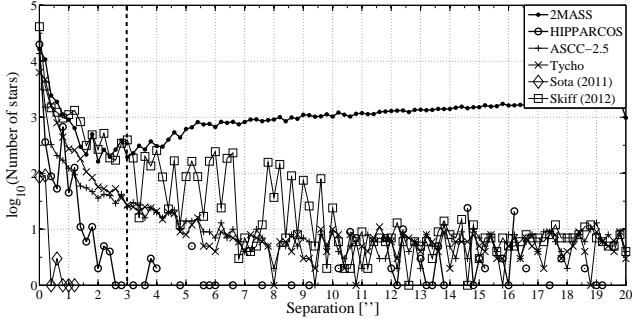}
\caption{Calculating the optimal aperture from the distribution of the separation of catalogue stars towards a sample star. The catalogues are 2MASS, the new reduction of HIPPARCOS, ASCC-2.5, Tycho, the massive star survey by \citet{Sota2011} and the catalogue from \citet{Skiff2013}. The optimal aperture is the first local minimum and is found to be 3\farcs{0} for HIPPARCOS and 2MASS, indicated by the dashed vertical line.}
\label{fig:KatalogAperturen}
\end{figure}

We use a cross correlation to re-identify stars in different catalogues. To avoid a misidentification of stars, we determine an optimal aperture from counting the spatial separation of the catalogue stars towards a sample star within a maximum aperture radius of 20\arcsec (\autoref{fig:KatalogAperturen}).
Given the location of the minimum in the separation distribution we consider an optimal aperture of 3\farcs{0} for 2MASS and HIPPARCOS coordinates and use this result for all other catalogues.
The star with the minimal spatial separation between sample and catalogue coordinates within the optimal aperture is identified to be the correct counterpart. 
If identifiers like DM, HD or HIP number are available, we check those as well. 
The numbers of identified sample stars in different catalogues are listed in \autoref{Tab-A:Katalogerkennungen}.
\begin{table}
\centering
\caption{Number of identified stars in different catalogues within a search aperture of 3\farcs{0}, the references are [1] \citet{Skrutskie2006}, [2] \citet{Kharchenko2009}, [3] \citet{Skiff2013}, [4] \citet{Perryman1997}, [5] \citet{vanLeeuwen2007}, [6] \citet{Zacharias2005}, [7] \citet{Schroeder2004}, [8] \citet{Hoeg1997}, [9] \citet{Zacharias2013}, [10] \citet{Bondarenko1996}, [11] \citet{Brancewicz1980}, [12] \citet{Docobo2006}, [13] \citet{Perevozkina1999}, [14] \citet{Pourbaix2004}, [15] \citet{Surkova2004}, [16] \citet{vanderHucht2001}, [17] \citet{Sota2011}, [18] \citet{Dommanget2002}, [19] \citet{Mason2010}}
\begin{tabular}{llc} \toprule
\textbf{catalogue name}                 & \textbf{stars} & \textbf{ref.}\\ \midrule
\textsc{2MASS}                          & 22\,755  & [1]\\
Allsky Comp. Cat. of 2.5 mio stars      & 19\,597  & [2]\\  
Cat. of Stellar Spectral Class.         & 9\,087   & [3]\\     
\textsc{Hipparcos} Main Cat.            & 6\,613   & [4]\\
\textsc{Hipparcos}, New Reduction       & 6\,613   & [5]\\
\textsc{NOMAD}                          & 23\,193  & [6]\\
Hipparcos parallaxes of O stars         & 154     & [7]\\
\textsc{Tycho} Main Part                & 18\,415  & [8]\\
\textsc{UCAC4}                          & 22\,387  & [9]\\
early-type contact binaries             & 31      & [10]\\
Parameters of eclipsing binaries        & 163     & [11]\\
Spec. subcomp. in mult. systems         & 4       & [12]\\
Cat. of eclipsing binaries parameters   & 9       & [13]\\
$9^{th}$ Cat. of Spec. Binary Orbits    & 531     & [14]\\
Semi-detached eclipsing binaries        & 37      & [15]\\
$7^{th}$ Cat. of Galactic WR stars      & 226     & [16]\\
Galactic O-Star Spectral Survey			& 184	  & [17]\\
CCDM                                    & 2\,506   & [18]\\
WDS                                     & 2\,595   & [19]\\ \hline
total number of different stars 					& 24\,948  & \\
\bottomrule
\end{tabular}
\label{Tab-A:Katalogerkennungen}
\end{table} 

\begin{figure}
\includegraphics[width=\columnwidth, trim=90 5 71 33,clip=true]{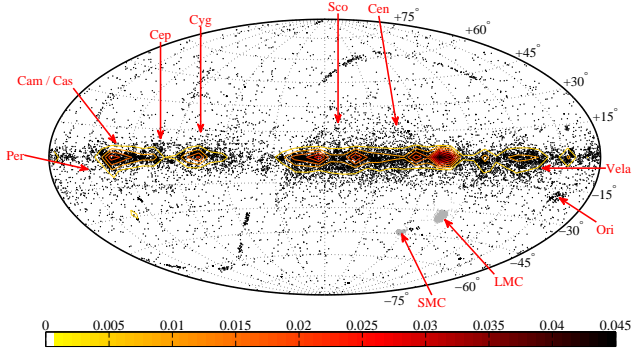}
\caption{The spatial distribution of our sample stars in galactic coordinates created with SIMBAD and various VizieR catalogues (see \autoref{sec:sample}), the grey dots indicate stars in the sky area of the Large and Small Magellanic Cloud which we exclude from our sample.}
\label{fig:spatialdistribution}
\end{figure}

The spatial distribution in galactic coordinates, of our whole sample stars is plotted in \autoref{fig:spatialdistribution}. 

\section{Stellar input parameters}
We collect for each star observational parameters i.e. the spectral type, BVRIJHK photometry, and the distance from the catalogues mentioned above (see \autoref{Tab:Allstars} for detailed list).
\begin{table*}
\centering
\caption{Stellar input parameters for ten progenitor stars of our sample, sorted by the ascending relative error of the parallaxes. The complete table with 24\,948 stars plus 79 resolved massive companions in multiples (\autoref{sec:multiple}) will be available at the CDS/ViZieR online data base. See Section 2 \& 3 for references.}
\begin{tabular}{clrrrrrrll}\toprule
\# & \multicolumn{1}{c}{ID} & \multicolumn{1}{c}{B} & \multicolumn{1}{c}{V} & \multicolumn{1}{c}{J} & \multicolumn{1}{c}{H} & \multicolumn{1}{c}{K} & \multicolumn{1}{c}{$\pi$} & \multicolumn{1}{c}{SpType} & \multicolumn{1}{c}{$T_{eff}$} \\
    &     & \multicolumn{1}{c}{[mas]} & \multicolumn{1}{c}{[mas]} & \multicolumn{1}{c}{[mas]} & \multicolumn{1}{c}{[mas]} & \multicolumn{1}{c}{[mas]} &\multicolumn{1}{c}{[mas]} & & \multicolumn{1}{c}{[K]}\\ \midrule
1 & $\alpha$ Mus &2.513(2) &2.681(2) &3.18$\pm$0.25 &3.17$\pm$0.21 &3.25$\pm$0.28 &10.44$\pm$0.11 &B2IV-V &$22\,000^{+3\,300}_{-3\,400}$ \\
2 & $\eta$ Cen &2.174(3) &2.328(4) &2.75$\pm$0.29 &2.78$\pm$0.21 &2.75$\pm$0.26 &10.78$\pm$0.21 &B2Ve &$22\,000^{+3\,300}_{-3\,400}$ \\
3 & $\beta$ Lup &2.493(2) &2.668(2) &3.15$\pm$0.21 &3.16$\pm$0.17 &3.25$\pm$0.27 &8.61$\pm$0.18 &B2IV &$21\,150^{+3\,250}_{-3\,550}$ \\
4 &HR 4898 &3.834(2) &3.993(2) &4.56$\pm$0.23 &4.56$\pm$0.08 &4.53$\pm$0.02 &7.95$\pm$0.17 &B1V &$25\,400^{+3\,400}_{-4\,600}$ \\
5 &  $\zeta$ Oph &2.598(3) &2.569(3) &2.53$\pm$0.30 &2.67$\pm$0.21 &2.68$\pm$0.27 &9.00$\pm$0.20 &O9.5IV &$27\,100^{+2\,400}_{-8\,150}$ \\
6 & $\alpha$ Lup &2.131(2) &2.282(2) &2.63$\pm$0.26 &2.72$\pm$0.20 &2.67$\pm$0.24 &7.09$\pm$0.17 &B2IV &$21\,150^{+3\,250}_{-3\,550}$ \\
7 &HR 5193 &3.280(3) &3.460(3) &3.71$\pm$0.21 &3.74$\pm$0.20 &4.01$\pm$0.04 &6.51$\pm$0.16 &B2Vnep &$22\,000^{+3\,300}_{-3\,400}$ \\
8 & $\phi$ Cen &3.618(3) &3.811(2) &4.63$\pm$0.28 &4.46$\pm$0.26 &4.49$\pm$0.02 &6.27$\pm$0.17 &B2V &$22\,000^{+3\,300}_{-3\,400}$ \\
9 & $\zeta$ Cas &3.490(2) &3.671(2) &4.14$\pm$0.30 &4.25$\pm$0.25 &4.25$\pm$0.04 &5.56$\pm$0.16 &B2IV &$21\,150^{+3\,250}_{-3\,550}$ \\
10 & $\upsilon$ Sco &2.502(4) &2.674(4) &3.05$\pm$0.23 &3.11$\pm$0.19 &3.18$\pm$0.26 &5.72$\pm$0.18 &B2IV &$21\,150^{+3\,250}_{-3\,550}$ \\
\bottomrule
\end{tabular}
\label{Tab:Allstars}
\end{table*}

\subsection{Temperature}
The effective temperature is calculated from the most recently published SpT available in the literature by using data from \citet{Lang1992}\footnote{taken from Schmidt-Kaler (1982)} and for main sequence stars from \citet{Kenyon1995}.
The temperature error is calculated from the given uncertainty in the SpT. 
If there is no given uncertainty we assume an error of the SpT by $\pm1$\,subclass. 
We interpolate the temperature between neighbouring grid points for a decimal SpT linearly.
For spectral types without a LC (3\,886 stars) we conservatively assume dwarf stars. If there is an uncertain LC (3\,800 stars), we consider the class with the lower luminosity (e.g. II/III is set to III). 
In total we collect full spectral classification for 24\,120 stars.

\subsection{Photometry and Extinction}
\begin{figure}
\includegraphics[width=\columnwidth, trim=195 80 170 51, clip=true]{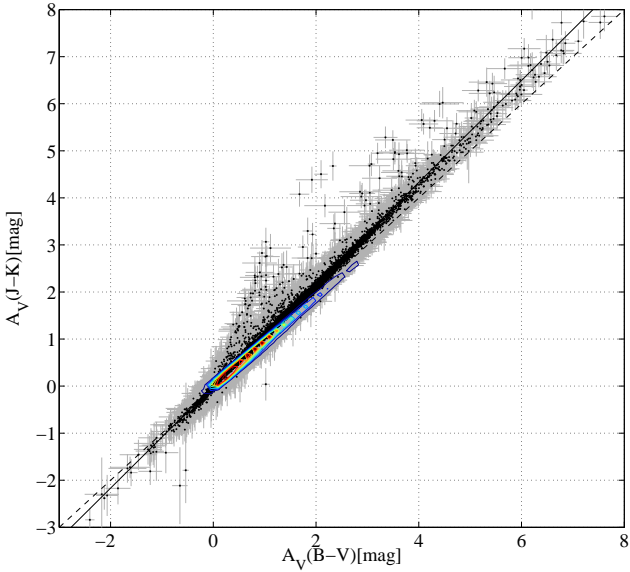}
\caption{Comparison of the derived $A_V$ using the colours B-V and J-K with 1$\sigma$ error bars. The one to one relation is indicated with a dashed line, the black line indicates the error weighted linear fit, and the coloured region shows the number density of stars from low (blue) to high (red). See \autoref{Tab:AVfit} for the results of the other colour combinations.}
\label{fig:AVplot}
\end{figure}

We use \textit{BVRIJHK} photometry with Johnson \textit{BV} from the HIPPARCOS and Tycho catalogues given in \citet{Kharchenko2009}, \textit{BV} from UCAC4 \citet{Zacharias2013}, the Cousins \textit{JHK} photometry from 2MASS, and \textit{BVRIJHK} from SIMBAD.
Having two different fluxes $X$ and $V$ the interstellar reddening $A_V$ due to interstellar extinction is calculated with the measured colour $(X-V)_{m}$ and the modelled colour $(X-V)_0$ according to:
\begin{equation}
\centering
A_V = \frac{(X-V)_{m} - (X-V)_{0}}{\frac{A_X}{A_V}-1}
\end{equation}
We compute $A_V$ for the colours (B-V), (V-J), (V-H), (V-K), (V-R), (B-I), (B-R), (R-I), (J-H), (J-K), and (V-I) with model colours from \citet{Bessell1998}. The model depends on $\log g$ and temperature, and hence on spectral classification.
As an example \autoref{fig:AVplot} shows the $A_V$ calculated from (J-K) and (B-V) as well as the number density of the values to emphasize the most frequent data points. The results of the error weighted linear fits are listed in \autoref{Tab:AVfit}.
The final $A_V$ is calculated as an error weighted mean using all the values from the different colours mentioned above. 
Since the constants $\frac{A_X}{A_V}$ published by \citet{Cardelli1989}, \citet{Rieke1985} and \citet{Savage1979} are slightly different, the mean of these values is used.

WR stars have typically very broad emission lines, so we use available Smith narrowband photometry and extinction $A_v$ given by \citet{vanderHucht2001} and calculate $A_V= 0.9149 \cdot A_v$ as recommended in their paper.
\begin{table}
\centering
\caption{The results of the error weighted fits $A_{V1}=m\cdot A_{V2}+n$, where $A_{V1}$ and $A_{V2}$ are interstellar extinction values derived from the named colours, $\bar{\sigma}_{A_{V1,2}}$ is the median of the respective error in mag, \# is the number of stars with an error smaller than 3\,$\bar{\sigma}_{A_{V1,2}}$ that are used for the fit. All results are consistent with a one-to-one relation and an offset of $n=0$.}
\small
\begin{tabular}{cccccc} \toprule
$A_{V1}$ & $A_{V2}$  & 3\,$\bar{\sigma}_{A_{V1}}$ & 3\,$\bar{\sigma}_{A_{V2}}$  & \#  & $m$ \\ \midrule
B-V &V-J &0.35 &0.22 &10\,399 &0.7$\pm$1.8 \\
B-V &V-H &0.35 &0.23 &10\,570 &0.9$\pm$1.8 \\
B-V &V-K &0.35 &0.17 &10\,618 &1.0$\pm$2.7 \\
B-V &J-H &0.35 &1.63 &10\,296 &1.1$\pm$0.3 \\
B-V &J-K &0.35 &0.85 &10\,315 &1.1$\pm$0.4 \\
V-J &V-H &0.22 &0.23 &11\,973 &1.2$\pm$1.6 \\
V-J &V-K &0.22 &0.17 &11\,921 &1.3$\pm$2.0 \\
V-J &J-H &0.22 &1.63 &11\,965 &1.5$\pm$1.5 \\
V-J &J-K &0.22 &0.85 &11\,987 &1.5$\pm$1.5 \\
V-H &V-K &0.23 &0.17 &12\,057 &1.1$\pm$1.6 \\
V-H &J-H &0.23 &1.63 &12\,131 &1.2$\pm$0.9 \\
V-H &J-K &0.23 &0.85 &12\,097 &1.2$\pm$1.0\\
V-K &J-H &0.17 &1.63 &11\,865 &1.1$\pm$1.2 \\
V-K &J-K &0.17 &0.85 &11\,923 &1.1$\pm$1.2 \\
\bottomrule
\end{tabular}
\label{Tab:AVfit}
\end{table}

\begin{figure}
\centering
\includegraphics[width=\columnwidth, trim=170 90 144 105, clip=true]{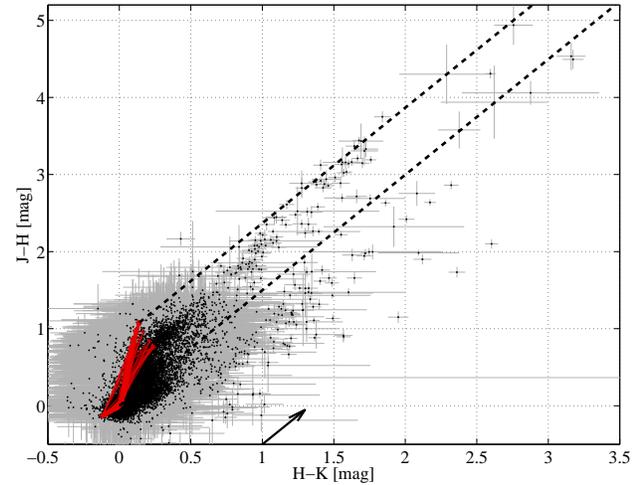}
\caption{The colour-colour diagram shows the measured (H-K) vs. (J-H) for our set of stars (black dots). The red lines are showing the modelled colours by \citet{Bessell1998} from dwarfs to hyper giants, the black arrow indicates interstellar extinction of 5\,mag, and the upper and lower dashed lines are indicating the reddening of stars with an effective temperature of 3\,500 and 50\,000\,K. Stars to the lower-right of the lower dashed line have circumstellar material.}
\label{Fig:JHKExzess}
\end{figure}

We distinguish between an excess due to circumstellar dust and interstellar reddening by using infrared colour-colour diagrams (e.g. \autoref{Fig:JHKExzess}).
There are 679 stars with an infrared excess due to circumstellar dust of more than 3$\sigma$ (colour error) in the JHK colour-colour diagram, which we take into account and correct for. Thus we avoid non-physical values of $A_V$.

\subsection{Multiplicity}
\label{sec:multiple}
Unresolved multiple stars would appear more luminous than resolved stars, hence the measured flux of unresolved multiple systems needs to be distributed among the components. 
Spectroscopic binaries are identified with catalogues from \citet{Brancewicz1980}, \citet{Bondarenko1996}, \citet{Pourbaix2004}, \citet{Perevozkina1999}, \citet{Dommanget2002}, \citet{Mason2010}, \citet{Docobo2006}, \citet{Surkova2004} and the SIMBAD object type (\autoref{Tab:Multikorr}).
We assume an identical interstellar extinction for unresolved members of a particular multiple and correct the apparent measured magnitudes for the components with:
\begin{eqnarray}
X_n&=&V_{n,m}-V_{sys,m}-\left(X-V\right)_{sys,0} \nonumber \\
& &+\left(X-V\right)_{n,0}+X_{sys,m}
\label{eq:multiplecorrection}
\end{eqnarray}
$X_{sys,m}$ indicates the measured magnitude of the system in the band X and $V_{n,0}$ is the magnitude in the V-band of the $n$-th component.
Thus we take into account the spectral classification and the measured magnitudes of the resolved components.

\begin{table}
\centering
\caption{Numbers of multiple systems found in our sample with different catalogues}
\begin{tabular}{ll} \toprule
      Catalogue               & multiple systems \\ \midrule
      CCDM                    & 1\,913 \\
      WDS                     & 2\, 844 \\
      \citet{Bondarenko1996}  & 33 \\
      \citet{Brancewicz1980}  &  175 \\
      \citet{Docobo2006}      & 5 \\
      \citet{Perevozkina1999} & 9 \\
      \citet{Pourbaix2004}    & 574 \\
      \citet{Surkova2004}     & 39 \\
\bottomrule
\end{tabular}
\label{Tab:Multikorr}
\end{table}

In our whole sample there are 4\,332 stars (17\% of our whole set) known to be members of 2\,131 multiple systems. This number is too small given the fact that 60-80\% of the massive stars are expected to be components of a multiple system \citep{Zinnecker2007}.
For the visually resolved multiple systems we list all resolved components with spectral classifications fulfilling our criteria from \autoref{tab:selcriteria} as additional entries in \autoref{Tab:Allstars}.

In total we count 24\,948 stars in \autoref{Tab:Allstars} plus 79 additional visual resolved massive companions.
There are 925 unresolved multiples with additional known components not fulfilling our selection criteria, and 1\,197 unresolved multiples where nothing else (except their existence) is known about the companions, which we assume to be binaries with the same spectral type.

\subsection{Distance}
\label{sec:distance}
The distance is calculated from the trigonometric parallax $\pi$ using the HIPPARCOS data of the recent reduction by \citet{vanLeeuwen2007}. 
If there is no HIPPARCOS value available for a star, the Tycho data or the catalogue from \citet{Kharchenko2009} is used.
The trigonometric parallax is available for 18\,300 sample stars.
Among them, we count 11\,862 stars with $\pi \geq 0.2\,$mas, 10\,102 stars with a value higher than its error $\vert \frac{\sigma_{\pi}}{\pi} \vert \leq 1$, and 6\,191 with a negative parallax. 
All parallaxes $\pi$ are transformed to the expectation value $\pi_t$ using Eq. 21 from \citet{Smith1996}.
The influence of this transformation on parallaxes with large errors is seen in \autoref{tab:plx}. There would be an overpopulation of 7\,945 stars with significantly underestimated distances resulting in inconsistent luminosities compared to their SpT and LC.

\begin{table}
\centering
\caption{Number of stars with measured parallax $\pi$ and transformed parallax $\pi_t$ higher than its error $\sigma$.}
\begin{tabular}{lcccc} \toprule
$\vert \frac{\pi}{\sigma_{\pi}} \vert$ & $\geq$\,5 & $\geq$\,3 & $\geq$\,2 & $\geq$\,1 \\ \midrule
measured $\pi$   & 1\,913 & 3\,281 & 5\,293  & 10\,102\\
transformed $\pi_t$ & 1\,954 & 3\,761 & 13\,883 & 18\,047\\
both $\pi$ and $\pi_t$   & 1\,912 & 3\,276 & 5\,264 & 9\,978\\
\bottomrule
\end{tabular}
\label{tab:plx}
\end{table}

\begin{figure}
\includegraphics[width=\columnwidth, trim=193 10 148 26, clip=true]{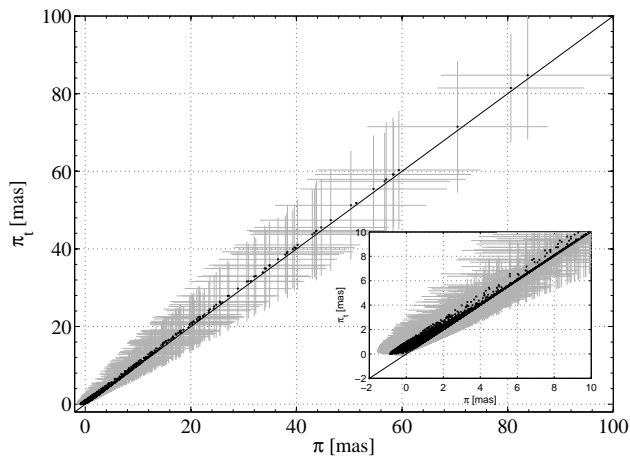}
\caption{Transformed parallax $\pi_t$ compared to the measured parallax $\pi$ using the transformation from \citet{Smith1996} for 6\,407 stars with $\vert \pi_{t}-\pi \vert \leq 1\,$mas. The solid line indicates the one-to-one relation. The interval up to 10\,mas is shown in the lower right.}
\label{fig:PlxTransformation}
\end{figure}

Since negative parallaxes are converted to small positive values within their errorbars, we flag those values as uncertain to exclude them from further calculations.
We use the transformation to utilize those parallaxes which are comparable to their expectation value and select 6\,407 stars with $\vert \pi_{t}-\pi \vert \leq \, 1\,$mas (\autoref{fig:PlxTransformation}).
Thus we avoid uncertain parallaxes with large errors but include measurements accurate down to sub-mas.

The results of the remaining part of the paper are calculated with and without an applied transformation to study systematic uncertainties that may arise from Eq. 21 by \citet{Smith1996}.
All values are consistent within their errorbars (see \autoref{sec:appendix}).

\subsection{Luminosity}
The bolometric luminosity in solar units is
\begin{equation}
L_{bol}=10^{0.4 \left( 5 \log_{10}d - 5 + 4.74mag - BC_V - m_V + A_V \right)}
\label{eqn:Lbol}
\end{equation}
with the distance $d$ in pc, the bolometric correction in the V-band \textbf{$BC_V$}, the apparent magnitude in the V-band $m_V$, and the interstellar extinction in the V-band $A_V$. The $BC_V$ from \citet{Bessell1998} is calibrated to $M_{bol \odot}=4.74$\,mag and is consistent with the apparent magnitude of $V_{\odot}=-26.76$\,mag as mentioned in their paper. This test is suggested by \citet{Torres2010}, because the zero point of the $BC_V$ is arbitrary, but the bolometric magnitude of the sun is not.
\begin{figure}
\includegraphics[width=\columnwidth, trim=186 33 154 41, clip=true]{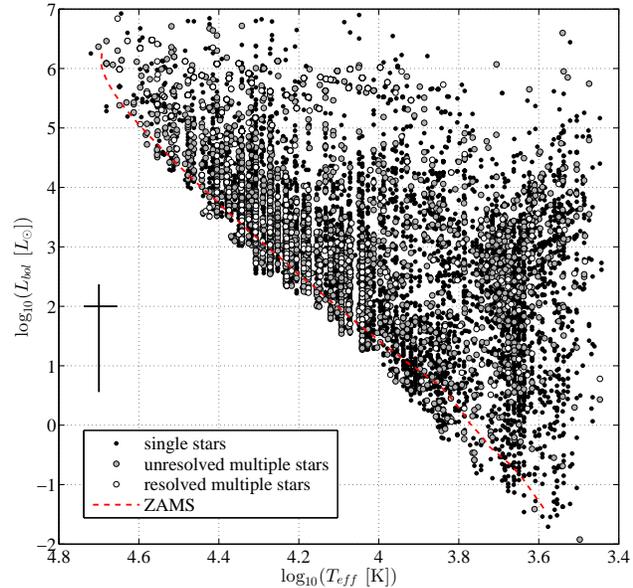}
\caption{Our derived H-R\,D: the red dashed line marks the zero age main sequence (ZAMS), grey dots indicate resolved stars in a multiple system, black circles denote unresolved multiples, and black dots stand for single stars, the mean errorbar is indicated by the black cross on the left side.}
\label{fig:HRDmulti}
\end{figure}

\section{Derived stellar parameters: mass, age, and life-time}
\label{sec:derivedparam}
We estimate mass and age by comparing our calculated bolometric luminosities and temperatures with the theoretical H-R\,D from \citet{Bertelli1994}, \citet{Schaller1992}, and \citet{Claret2004}. 
The differences of these models among each other are described in \citet{Hohle2010}.
Our mass limit of 8\,M$_{\odot}$ corresponds to a main sequence star with $\log {\left(L_{bol}/L_{\odot}\right)} \geq$\,3.41 and $\log {\left( T_{eff}/[K] \right)} \geq$\,4.23 regarding the model from \citet{Claret2004}.
Mass and age are calculated from the error-weighted closest grid-points per model. We then calculate the mean mass and mean age from the different models.
For stars with $M \geq 35\,M_{\odot}$ we use the models from \citet{Claret2004} and \citet{Schaller1992}. The theoretical H-R\,D from \citet{Bertelli1994}, \citet{Schaller1992}, and \citet{Claret2004} are applied for stars with $M < 35\,M_{\odot}$. For WR stars we only use the model from \citet{Meynet2003}.

There are 8\,143 stars below the zero-age-main-sequence (ZAMS) due to large uncertainties of the distance, which are conservatively assumed to have at least a luminosity consistent with the ZAMS (as seen in \autoref{fig:HRDmulti}). So far we cannot distinguish between stars with underestimated distances and misidentified subdwarfs that also lie below the ZAMS, thus we flag those stars. From our 6\,407 stars from the subsample with sufficiently measured distances (\autoref{sec:distance}), we correct 2\,663 stars to be on the ZAMS. Those stars probably have erroneous distances which are not consistent to be within 1\,kpc, however they should have distances smaller than 5\,kpc due to the completeness of the catalogues for the parallax and the apparent magnitudes.

Since the spectroscopic binary catalogues from \autoref{Tab:Multikorr} list dynamical masses, we use these instead of model dependent masses if available. For main sequence stars those dynamical masses are consistent with our masses estimated from the H-R\,D within their 1\,$\sigma$ errorbars \citep{Docobo2013}. 

\begin{figure}
\includegraphics[width=\columnwidth, trim=147 150 96 124, clip=true]{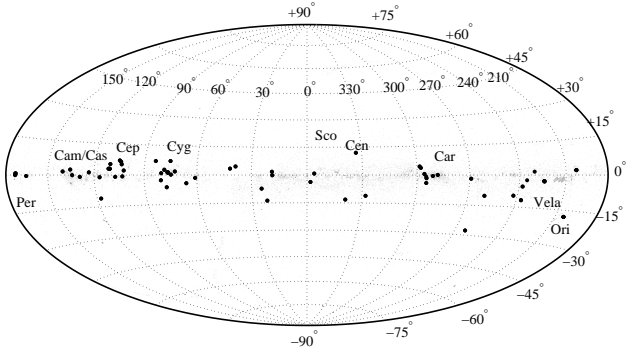}
\caption{The spatial distribution of 146 binaries, where both components have at least 8\,$M_{\odot}$ (black dots) and the entire sample (grey dots).}
\label{fig:galacticmassivebinaries}
\end{figure}

Massive multiple systems where at least two components have masses above 8\,$M_{\odot}$ are important for the study of the evolution of compact binary systems. The spatial distribution of those massive binaries is shown in \autoref{fig:galacticmassivebinaries} (listed in \autoref{Tab-A:MassereicheBinaries}).
In the near future, we will also study the binary evolution and interaction in those systems.

\begin{figure}
\includegraphics[width=\columnwidth, trim=160 79 189 130,clip=true]{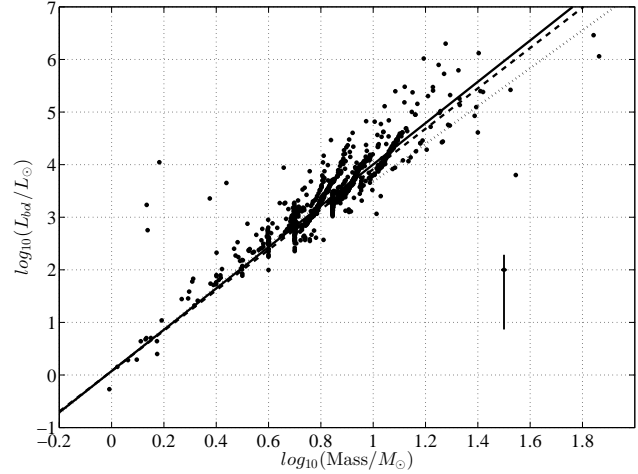}
\caption{Derived error weighted mass luminosity relation $L \sim M^{\beta}$ with $\beta=3.93\pm0.78$ (solid line) for main sequence stars compared to published values from \citet{Hilditch2001} with $\beta=4.0$ for stars with $M \leq 10\,M_{\odot}$ and $\beta=3.6$ for stars with a higher mass (dotted line), and $\beta=3.84$ from \citet{Andersen1991} (dashed line). The black cross indicates the mean errorbar.}
\label{fig:MassLum}
\end{figure}

The empirical mass luminosity relation $L \sim M^{\beta}$ yields $\beta=3.93\pm0.78$ for main sequence stars from our sample and is plotted in \autoref{fig:MassLum}. That error weighted result is consistent with the values from \citet{Hilditch2001} with $\beta=4.0$ for stars with a mass less than $10\,M_{\odot}$ and $\beta=3.6$ for stars with a higher mass, from \citet{Andersen1991} with $\beta=3.84$, and with $\beta=3.66\pm0.12$ from \citet{Hohle2010}.

\begin{figure}
\includegraphics[width=\columnwidth, trim=179 190 211 90,clip=true]{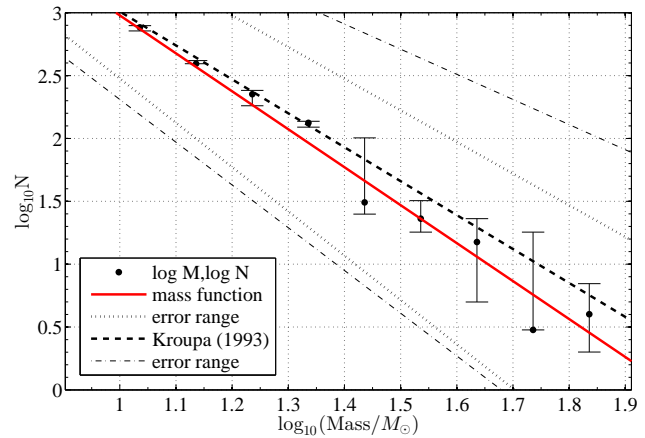}
\caption{Our derived mass function $N \sim M^\Gamma$ (red solid line) with $\Gamma=-3.1\pm0.5$ for stars with $M \gtrsim 8\,M_{\odot}$ compared to the initial mass function from \citet{Kroupa1993} with $\Gamma=-2.7\pm0.7$ (black dashed line). The errorbars of the individual masses (black dots) indicate the 1$\sigma$ confidence interval.}
\label{fig:MassFunc}
\end{figure}

As another consistency check we calculate the mass function $N \sim M^\Gamma$ for massive stars by using an error weighted fit yielding $\Gamma=-3.1\pm0.5$ (see \autoref{fig:MassFunc}), that is in good agreement with $\Gamma=-2.7\pm0.7$ from the initial mass function by \citet{Kroupa1993}.

The H- and He-burning phases specify the life-time of a star due to the long duration (Myr) while other burning processes like C-burning lasts a few kyr only. The life-time of a star can be calculated from its mass by using models of \citet{Kodama1997} and \citet{Maeder1989}.
The remaining life-time $\tau_r$ of a star is the difference between the mean of the expected modelled life-time $\tau_{m}$ and the age $\tau_{age}$.
\begin{equation}
\tau_r=\tau_{m}-\tau_{age}
\end{equation}

Since $\tau_{m}$ is not well calibrated for the theoretical H-R\,D, there are 9\,563 stars in our sample with a negative remaining life-time.
All stars with a negative remaining life-time are randomly set to plausible life-times according to the distribution of positive life-times of their respective mass bin. Hence, the remaining life-time distribution of every mass bin itself is not changed by this correction.
The corrected life-times are marked with a leading colon in \autoref{Tab:DerivedParameters}. If there is no positive life-time distribution for a mass bin, we flag those life-times with a question mark. We count 2\,643 stars with a corrected remaining life-time from the subsample of 6\,407 stars with sufficient distances (\autoref{sec:distance}).

The calculated $L_{bol}$, the mass, the age, the remaining life-time $\tau_r$, and the number N of known additional components are listed in \autoref{Tab:DerivedParameters}.

\begin{table}
\centering
\caption{List of $L_{bol}$, mass, age, remaining life-time $\tau_r$, and number N of known additional components of the first ten SN progenitors in \autoref{Tab:Allstars}; with 1\,$\sigma$ errors. Uncertain remaining life-times are marked with a leading colon. The complete table with 25\,027 entries will be published on the CDS/ViZieR online data base.}
\begin{tabular}{llrrrrl} \toprule
\# & name              & $L_{bol}$      & mass & age & $\tau_r$ & N \\ 
    &                       & $10^3$ [$L_{\odot}$]  & [$M_{\odot}$] & [Myr] & [Myr] & \\ \midrule   
    1 & $\alpha$ Mus &5.3$\pm$2.4 &8.5(2) &15$\pm$3 &14 &4 \\
    2 & $\eta$ Cen &7.2$\pm$3.4 &8.7(2) &20$\pm$3 &8 &0 \\
    3 & $\beta$ Lup &6.7$\pm$3.1 &8.2(3) &28$\pm$6 &:12 &1 \\
    4 &HR 4898 &3.7$\pm$1.7 &8.9(1) &3$\pm$1 &24 &1 \\
    5 &$\zeta$ Oph &21.9$\pm$2.5 &12.5(1) &13$\pm$1 &2 &0 \\
    6 & $\alpha$ Lup &15.4$\pm$7.4 &9.6(2) &23$\pm$1 &:3 &2 \\
    7 &HR 5193 &6.7$\pm$3.3 &8.7(2) &19$\pm$4 &9 &2 \\
    8 & $\phi$ Cen &4.6$\pm$2.3 &8.3(2) &14$\pm$2 &17 &0 \\
    9 &$\zeta$ Cas &6.2$\pm$3.1 &8.1(2) &28$\pm$5 &4 &0 \\
    10 &$\upsilon$ Sco &16.1$\pm$8.0 &9.7(3) &23$\pm$1 &:7 &0 \\
\bottomrule
\end{tabular}
\label{Tab:DerivedParameters}
\end{table}

\subsection{Completeness}
\begin{figure}
\includegraphics[width=\columnwidth, trim=300 175 286 150, clip=true]{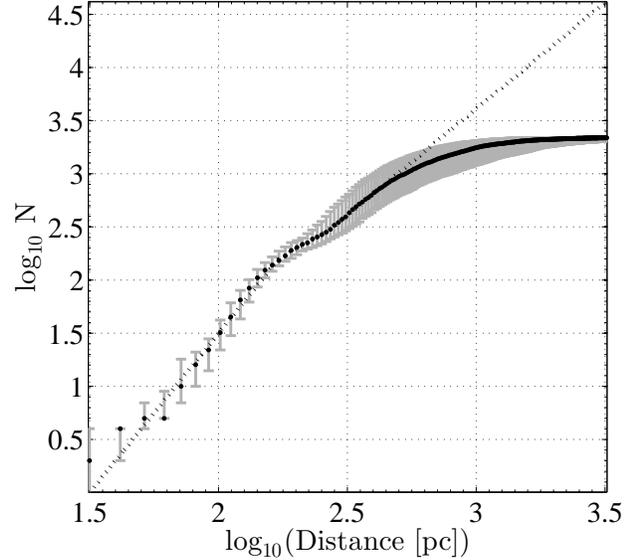}
\caption{The cumulative number of stars $N$ (black dots with grey errorbars) versus the distance $d$; our sample is consistent with a powerlaw (black dotted line) $N \sim d^{\Phi}$ of $\Phi=$3.0 for distances between 1 to 170\,pc, $\Phi=$1.4 from 170\,pc to 260\,pc, and $\Phi=$2.0 for 260\,pc to 500\,pc. Thus we estimate the completeness of our sample with the powerlaw index of 2.0 compared to the cumulative number. For a distance of 500\,pc our sample is complete to 97$\%$, 85$\%$ for 600\,pc, 45$\%$ for 1\,kpc, 6$\%$ for 3\,kpc, and 2$\%$ for 5\,kpc. The grey errorbars are obtained by adding or subtracting 1$\sigma$ from the parallax.}
\label{fig:completeness}
\end{figure}

The substantial completeness of our sample is conservatively calculated in \autoref{fig:completeness} by dividing the cumulative number of stars at a certain distance by the value from a fitted power-law growth $N\sim\,d^{\Phi}$. 
The scale height of the spiral arm population is $\approx\,$120\,pc and the young disc population has $\approx\,$200\,pc in \citet{Mihalas1981}. Our sample corresponds to a cubic growth ($\Phi=$3) for distances from 1 to 170\,pc as expected for increasing the observed spheric volume and counting the star numbers. Thus, the scale height of our stars is 170\,pc.
The scale height of the Galactic disc yields 260\,pc \citep{Harding2001}. The best fit is $\Phi=$1.4 from 170\,pc to 260\,pc, which agrees to a less dense populated region up to the scale height. Corresponding to an increase of the observed area within the Galactic disc a quadratic growth ($\Phi=$2) fits for distances from 260\,pc to 500\,pc. After 500\,pc the incompleteness of our sample increases, hence we cannot longer fit a powerlaw growth to the cumulative number.
Dividing the cumulative number by the value from the continued quadratic powerlaw for distances higher than 500\,pc our set of stars is 85\,$\%$ complete within 600\,pc, 45\,$\%$ within 1\,kpc, 6\,$\%$ within 3\,kpc, and 2\,$\%$ within 5\,kpc. Thus, for distances larger than 1\,kpc our sample is highly incomplete.

\section{Discussion on Supernova progenitor distribution}
If the current local SN rate is roughly constant for some 10 to 20 Myr, then all ccSN progenitors with a mass higher than 8\,$M_{\odot}$ and a remaining life-time shorter than 10\,Myr are indicators for regions with young neutron stars and potential GW sources.
We can discuss our observed set of stars only within the next 10\,Myr because afterwards new star formation must be taken into account e.g. by performing a population synthesis.

\begin{figure}
\includegraphics[width=\columnwidth, trim=125 116 84 131,clip=true]{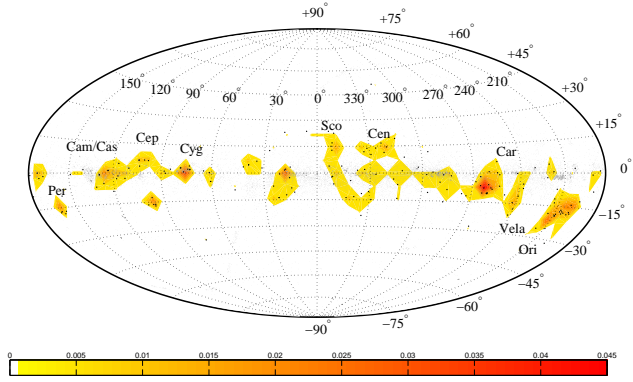}
\caption{The spatial distribution of regions with a local increased SN rate within 600\,pc from the sun, for the next 10\,Myr; all SN events (black dots) will occur in 8.6\,\% of the whole sky and 90\,\% of the SN progenitors are distributed among 7.2\,\% of the whole sky. The bin size is 7\,$^\circ\times$7\,$^\circ$.}
\label{fig:snrate600pc}
\end{figure}

The regions with a locally increased SN rate within 600\,pc are shown of the whole sky with a binsize of 7$^\circ\times$7\,$^\circ$ in \autoref{fig:snrate600pc}. The SN rate is increased for OB clusters like Carina, Orion, Vela, Scorpius, Centaurus, Cygnus, Cepheus, and Perseus, which all belong to the Gould Belt. All SN progenitors are distributed among 8.6\,$\%$ \textbf{area} of the whole sky and 90\,$\%$ of the progenitors are within 7.2\,$\%$ area of the whole sky. 

\begin{figure}
\includegraphics[width=\columnwidth, trim=127 107 86 127,clip=true]{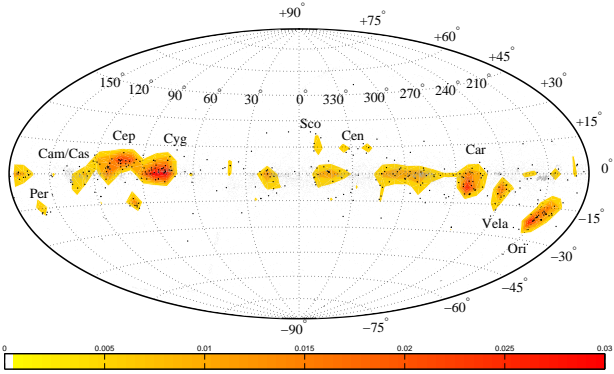}
\caption{The spatial distribution of regions with a local increased SN rate within 1\,kpc for the next 10\,Myr; all SN progenitors (black dots) are distributed among 12.2\,\% area of the whole sky and 90\,\% of the SN events are in 9.5\,\% of the whole sky. The bin size is 7\,$^\circ\times$7\,$^\circ$.}
\label{fig:snrate1kpc}
\end{figure}

Within 1\,kpc (\autoref{fig:snrate1kpc}) all SN events will occur in 12.2\,\% of the whole sky and 90\,\% of the SNe are in 9.5\,\% of the whole sky.

\begin{figure}
\includegraphics[width=\columnwidth, trim=122 94 87 115,clip=true]{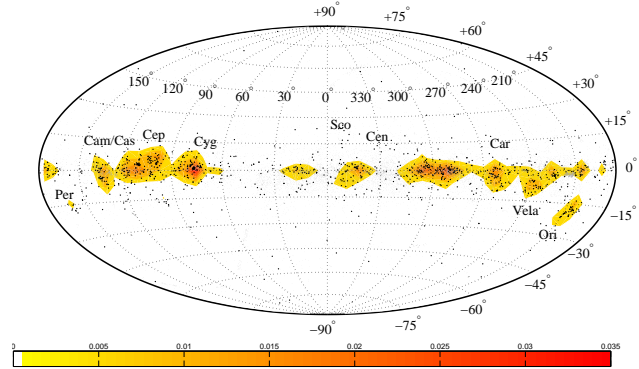}
\caption{The spatial distribution of regions with local increased SN rate within 5\,kpc for the next 10\,Myr; all SN events (black dots) in 21.0\% of the whole sky and 90\,\% SNe in 12.1\,\% of the whole sky. The bin size is 7\,$^\circ\times$7\,$^\circ$.}
\label{fig:snrate5kpc}
\end{figure}

The regions with a locally increased SN rate within 5\,kpc from the sun are plotted in \autoref{fig:snrate5kpc}. All SN progenitors are distributed among 21.0\,\% area of the whole sky and 90\,\% of the SN events will occur in 12.2\,\% area of the whole sky.

While our sample is incomplete outside $\approx$1\,kpc, the missing SN progenitors are possibly mostly in the same areas as the known ones.
These results can be used e.g. for reducing the parameter space of Einstein@home, because GW sources or young neutron stars should be concentrated in our marked regions, even when taking into account high kick velocities \citep{Palomba2005}. Thus a blind search for these sources could be reduced to our regions.

\begin{figure}
\includegraphics[width=\columnwidth, trim=141 78 100 80,clip=true]{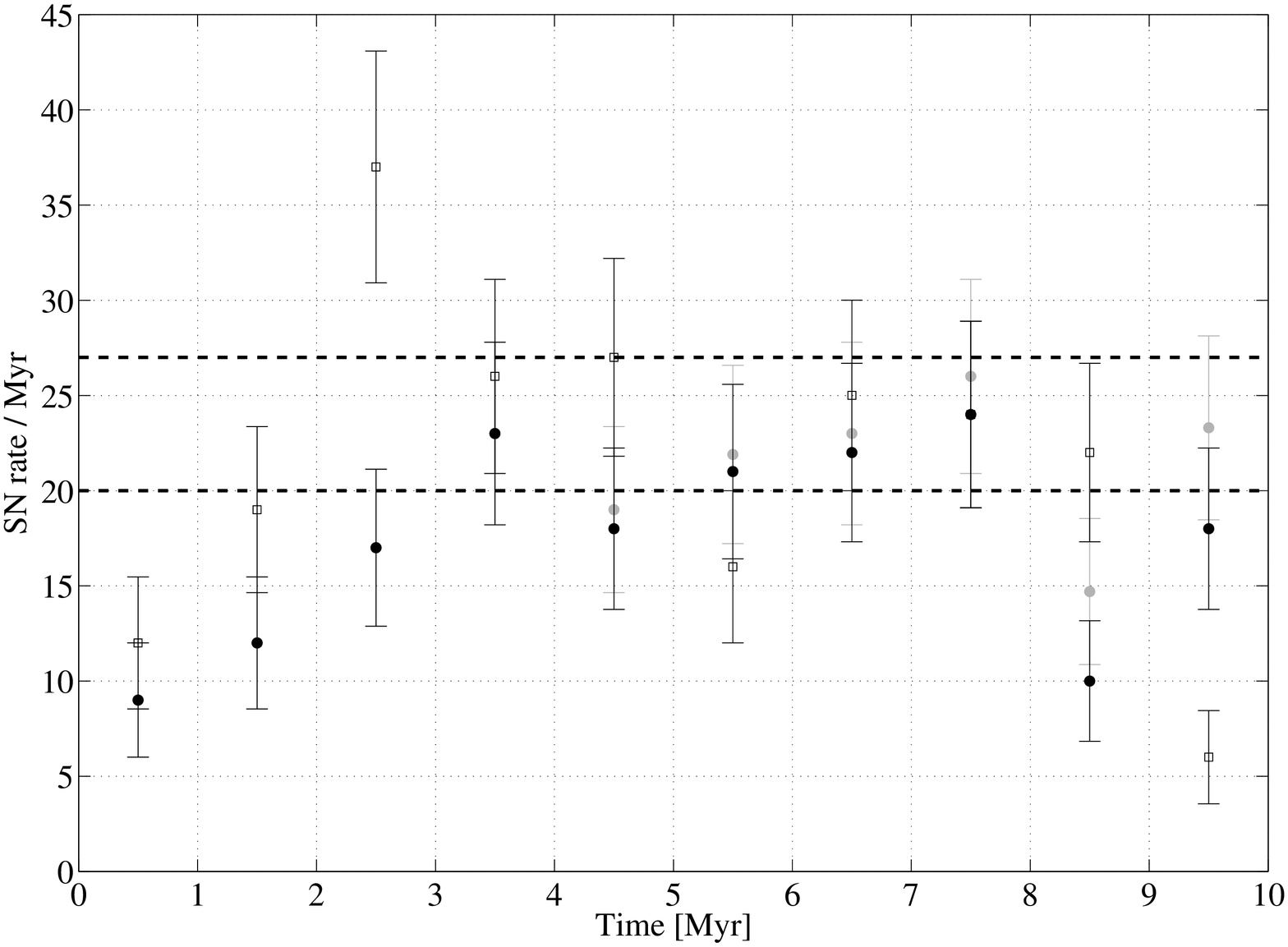}
\caption{The ccSN rate of $17.0^{+4.5}_{-3.6}$\,ccSNe/Myr from this work for the next 10\,Myr within 600\,pc (black filled dots with Poissonian errorbars) in comparison with the results in \citet{Hohle2010} (black open squares). The dashed line indicates the apparent range for the Gould Belt from \citet{Grenier2004}. The increase of the SN rate due to putative massive companions of possible multiple stars is shown by grey dots.}
\label{fig:snratemyr}
\end{figure}

The temporally resolved ccSN rate for the next 10\,Myr within 600\,pc yields $17.0^{+4.5}_{-3.6}$\,ccSNe/Myr and is shown in \autoref{fig:snratemyr}. Our result can be compared to the range from 20 to 27\,ccSNe per Myr for the Gould Belt from \citet{Grenier2004} and also with $21\pm5$\,ccSNe/Myr from \citet{Hohle2010}. The value between 2 and 3 Myr, seen as a peak in \citet{Hohle2010}, is now consistent with a constant rate. Within 2 Myr the SN rate seems to be slightly below the average.

The true SN rate can be larger due to SNe of type \,Ia, ($\approx$15-25\,\% of all SNe) plus more unknown multiple massive companions. For these stars we take into account the probability of putative $8\,M_{\odot}$ companions with the same age as the host star by using the binary fraction from \citet{Oudmaijer2010}. This slight increase is illustrated by the grey dots in \autoref{fig:snratemyr}.

Many neutron stars are expected to be nearby the Galactic Center, but there is no major contribution to the spatial distribution in that direction due to high uncertainties in the parallax and high extinction values.
The accurate photometric and astrometric results of the \textit{Gaia} mission \citep{deBruijne2012} could improve the completeness and the precision of our work significantly.

We consider stars with a mass higher than 8\,$M_{\odot}$ as SN progenitors and with a mass higher than 30\,$M_{\odot}$ as probable black hole (BH) progenitors \citep{Woosley2002}.
In \autoref{Tab:SNrate} we give the derived ccSN rate, the numbers of SN progenitors, and BH progenitors within a certain distance. We calculate the Galactic ccSN rate by extrapolating the local ccSN rate of our observed volume to the whole Galaxy as described in \citet{Reed2005} using his Equations (5) and (16). We use as distance to the Galactic Center 8.5\,kpc, a thick disc scale height of 0.9\,kpc, and a scale length of 3.6\,kpc \citep{Juric2008}.
The observed volume is then calculated within the disc using the distances given in \autoref{Tab:SNrate}. Comparing the ccSN rate in the Solar vicinity with the Galactic ccSN rate of 2.5 SNe per century \citep{Tammann1994} suggests that the current ccSN rate within 600\,pc and 1\,kpc is increased due to the Gould Belt as found by \citet{Grenier2004} and \citet{Popov2005} by about a factor of 5-6.

\begin{table*}
\centering
\caption{The number of SN progenitors with $M>8\,M_{\odot}$ and BH progenitors with $M>30\,M_{\odot}$ within a specific distance. The lower number is obtained by subtracting the 1$\sigma$ error from the masses and the upper value is obtained by adding the 1$\sigma$ error to the masses. The ccSN rate is given for those progenitors which should explode within the next 10\,Myr, the errors are 3$\sigma$. In this table we give the calculated ccSN rate for our Galaxy by extrapolating from the local SN rate (given in that row) to the whole Galaxy, with a distance to the Galactic Center of 8.5\,kpc, a thick disc scale height of 0.9\,kpc, and a scale length of 3.6\,kpc \citep{Juric2008}. In the last column we give the completeness from \autoref{fig:completeness}.}
\begin{tabular}{llllllll} \toprule
\textbf{distance} & \textbf{SN} & \textbf{BH} & \textbf{SN rate} & \textbf{Gal.SN rate}& \textbf{Completeness}\\ 
		$[$kpc$]$ & [\# progenitors] & [\# progenitors] & [SN/Myr] & [SN/100yr] & [\%]   \\ \midrule
        $\leq$\,0.6 & $273^{+27}_{-28}$ & $0\pm0$ & $16.9^{+4.5}_{-3.6}$ & $11.9^{+1.9}_{-0.9}$ &85  \\ \hline
		$\leq$\,1.0 & $518^{+44}_{-47}$ & $2\pm0$ & $34.8^{+6.0}_{-6.9}$ & $6.0^{+0.3}_{-0.4}$ & 45 \\ \hline
		$\leq$\,3.0 & $763^{+62}_{-63}$ & $6\pm1$ & $57.6^{+7.8}_{-9.6}$ &  $0.7^{+0.1}_{-0.1}$ & 6  \\ \hline
		$\leq$\,5.0 & $1\,328^{+78}_{-111}$ & $58^{+4}_{-7}$ & $110.2^{+9.0}_{-15.6}$ &  $0.4^{+0.1}_{-0.1}$ & 2    \\
\bottomrule
\end{tabular}
\label{Tab:SNrate}
\end{table*}

\section{Conclusions}
The sample contains one order of magnitude more stars than \citet{Hohle2010}, is more complete, and includes more recent input parameters. Our consistency checks for the mass luminosity relation, the mass function, and the completeness confirmed known relations and published values from different authors. We found regions with a locally increased ccSN rate implying a higher probability of the presence of young neutron stars.
The blind search for young nearby neutron stars and GW sources can work more efficient, if the parameter space of an all sky survey can be reduced to our predicted areas. 
Given the results of this work a blind search for GW sources and young neutron stars within 600\,pc could be decreased to only 9\,\% area of the whole sky. Regarding our highly incomplete sample within 5\,kpc, the all sky survey could be restricted to 21\,\% area of the whole sky.
The discovery of new radio quiet neutron stars, like in \citet{Pires2009} in Carina, which shows an enhanced probability in our maps, is more probable in these special regions than in a blind search.

\section*{Acknowledgements}
The authors would like to thank S. B. Popov, R. Chini, and V. V. Hambaryan for useful discussions,
the Deutsche Forschungsgemeinschaft (DFG) for support in the Collaborative Research Center Sonderforschungsbereich Transregio 7 "Gravitationswellenastronomie", Bruce Allen \& Bernard Schutz for motivating and suggesting this study, and Nina Tetzlaff for helping with plotting issues.\\
This research has made use of the SIMBAD database; data products from the Two Micron All Sky Survey, which is a joint project of the University of Massachusetts and the Infrared Processing and Analysis Center/California Institute of Technology, funded by the National Aeronautics and Space Administration and the National Science Foundation; the Washington Double Star Catalog maintained at the U.S. Naval Observatory.\\

\appendix

\section{Selection of the stars}
The basic set of stars comes from a \textit{select by criteria} query in SIMBAD using the following search expression:
\label{sel:SIMBAD}
\begin{equation}
\begin{array}{l}
(\left(\texttt{splum>=Ia0} \:\&\: \texttt{splum<=IIb} \right) \:\vert\: \\
\left( \left( \texttt{splum>IIb \& splum<IV} \right) \:\&\: \texttt{sptype<=B9}\right) \:\vert\: \\
\left( \texttt{sptype<=B4} \right) ) \:\&\:  \\
\left(\texttt{otypes='**'} \:\vert\: \texttt{otype='*'} \right)  \:\&\: \left( \texttt{ra<=360.0} \right)
\end{array}
\label{eqn:simbadexpression}
\end{equation}
where \texttt{splum} defines the spectral LC, \texttt{sptype} specifies the spectral type, \texttt{otype} is for the classification of the objects e.g. as stars or multiple stars, and the \texttt{ra} selection is to query stars with coordinates. 
The last argument in \autoref{eqn:simbadexpression} filters out 410 stars without listed coordinates. 
We expand the sample from SIMBAD by querying catalogues containing spectral classifications in the order of descending priority:
\citet{Sota2011}, CSSC by \citet{Skiff2013}, ASCC-2.5 by \citet{Kharchenko2009}, HIPPARCOS by \citet{Perryman1997}. The VizieR Online Database \citep{Ochsenbein2000} was used to download these catalogues.

\section{Calculations and Results without transformed Parallax}
\label{sec:appendix}
In this section we give all results derived without applying Eq. (21) from \citet{Smith1996}. The values are consistent within their errorbars compared to the values in the main part of the paper.

\begin{figure}
\includegraphics[width=\columnwidth, trim=197 46 149 35, clip=true]{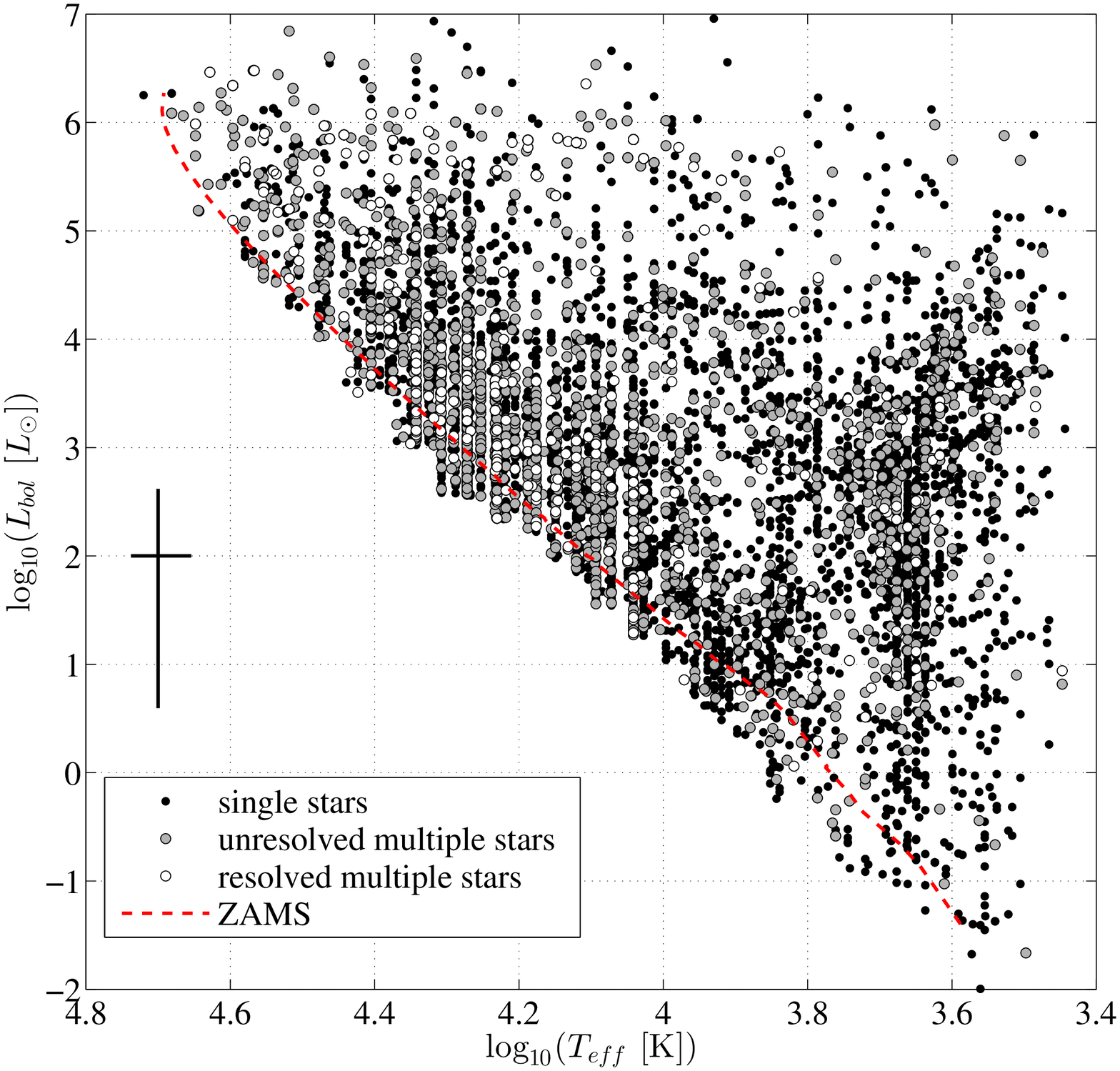}
\caption{Our derived H-R D: the red dashed line marks the zero age main sequence (ZAMS), grey dots indicate resolved stars in a multiple system, black circles are unresolved multiples, and black dots identify single stars, the mean errorbars are indicated by the black lines on the left side.}
\label{fig:HRDmulti_ohneSE}
\end{figure}

The calculated H-R\,D using spectral classification and \autoref{eqn:Lbol} is seen in  \autoref{fig:HRDmulti_ohneSE}.
There are in total 4~504 stars below the zero-age-main-sequence (ZAMS) due to large uncertainties of the distance. Those stars are flagged and assumed to have at least a luminosity consistent with the ZAMS. 
From our 6~407 stars with sufficiently measured distances (\autoref{sec:distance}), 518 stars are below the ZAMS.

\begin{figure}
\includegraphics[width=\columnwidth, trim=173 81 173 130,clip=true]{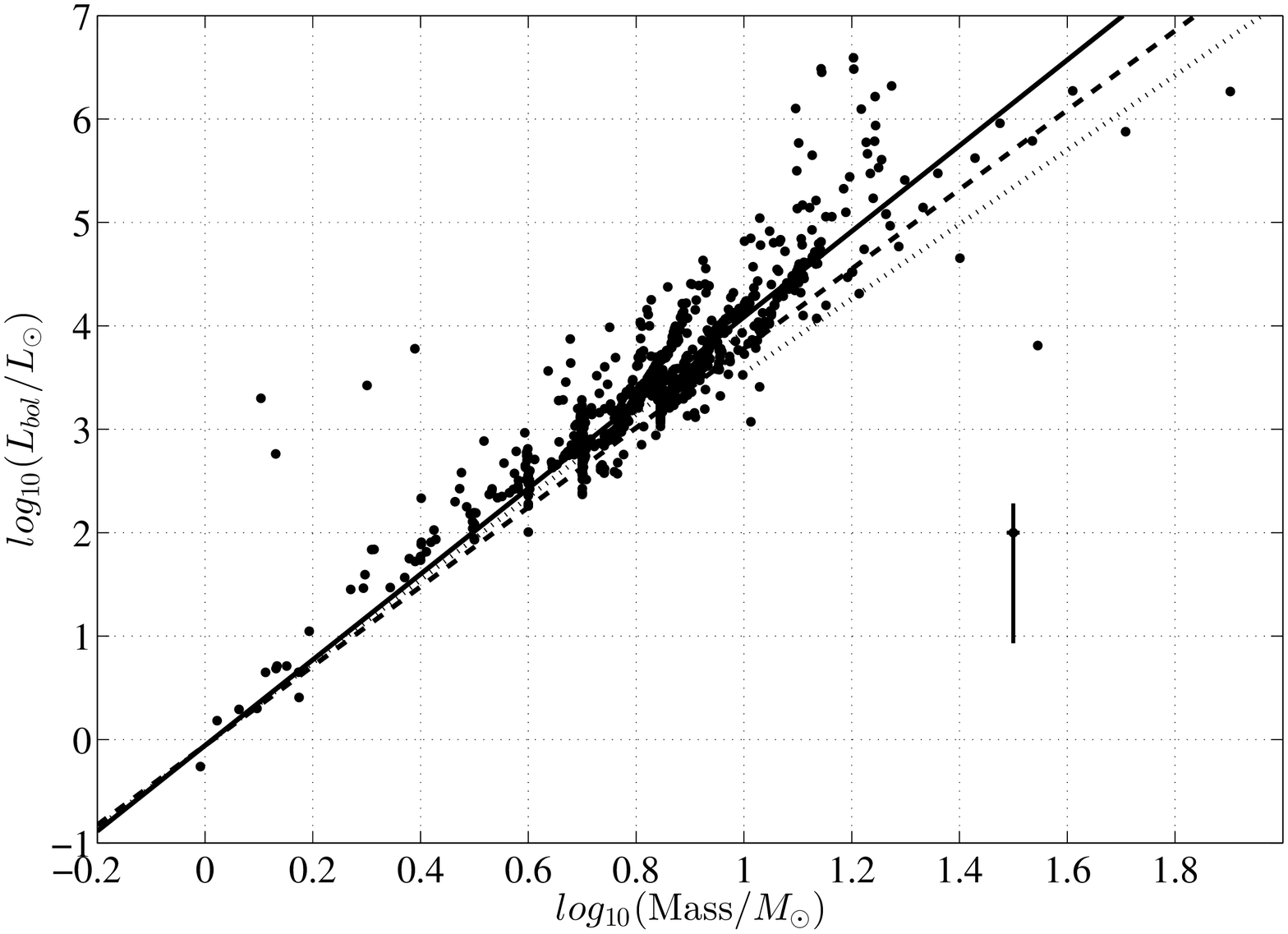} 
\caption{Derived error weighted mass luminosity relation $L \sim M^{\beta}$ with $\beta=4.14\pm0.92$ (solid line) for main sequence stars compared to with $\beta=$4.0 for stars with $M \leq 10\,M_{\odot}$ and 3.6 for stars with a higher mass (dotted line) from \citet{Hilditch2001}, and $\beta=3.84$ from \citet{Andersen1991} (dashed line). Our result is consistent with $\beta=3.66\pm0.12$ from \citet{Hohle2010}. The black line indicates the mean errorbar.}
\label{fig:MassLum_ohneSE}
\end{figure}

\begin{figure}
\includegraphics[width=\columnwidth, trim=276 111 160 132,clip=true]{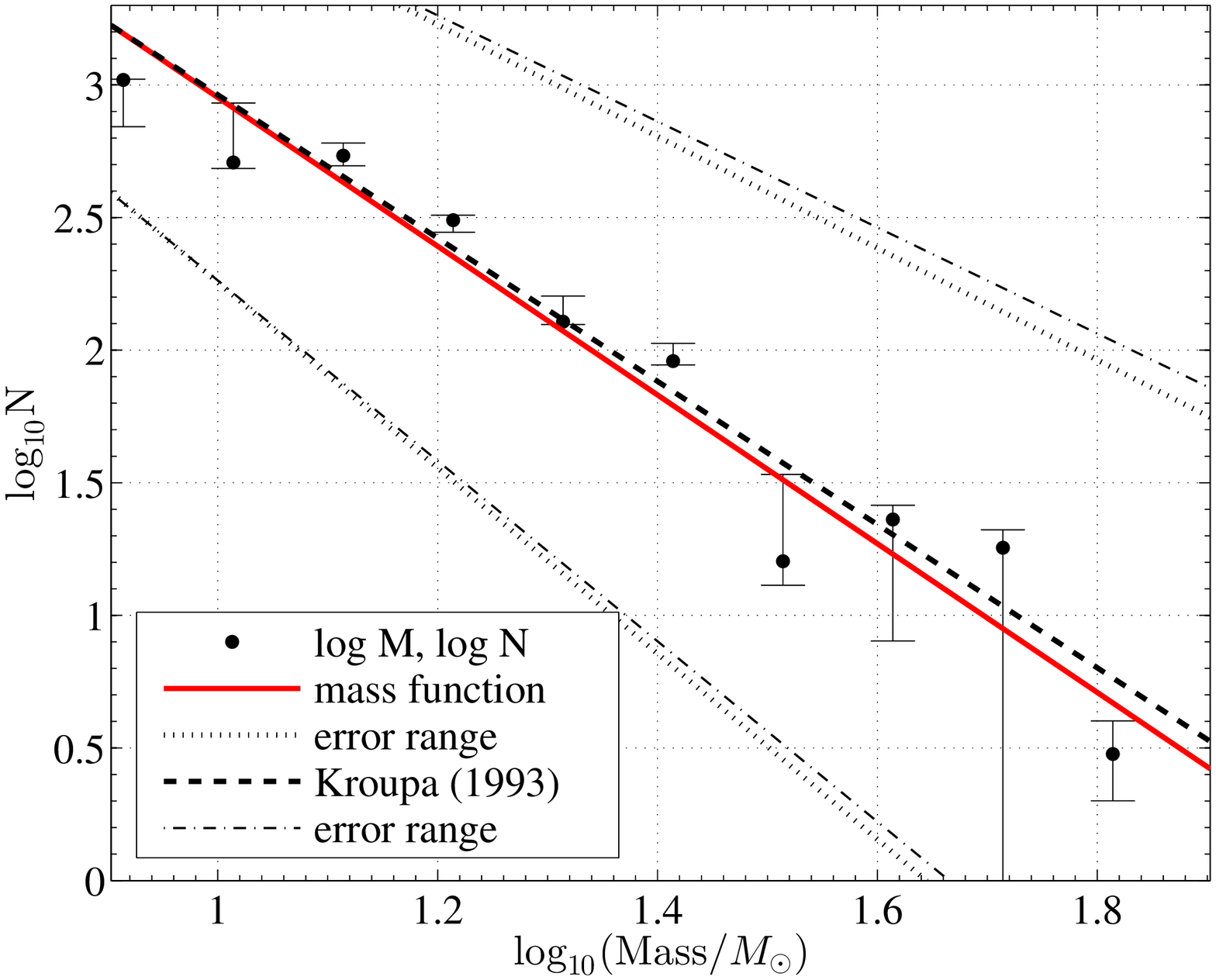}
\caption{Our derived mass function $N \sim M^\Gamma$ with $\Gamma=-2.8\pm0.7$ for stars with $M \gtrsim 8\,M_{\odot}$ (red line), wihout applied transformation by \citet{Smith1996}, compared to the initial mass function from \citet{Kroupa1993} with $\Gamma=-2.7\pm0.7$ (dashed black line). The errorbars of the individual masses (black dots) indicate the 1$\sigma$ confidence interval.}
\label{fig:MassFunction_ohneSE}
\end{figure}

The error-weighted fit for the mass luminosity relation of main sequence stars (\autoref{fig:MassLum_ohneSE}) yields $\beta=4.14\pm0.92$ and is consistent with $\beta=$4.0 for stars with $M \leq 10\,M_{\odot}$ and $\beta=$3.6 for stars with a higher mass from \citet{Hilditch2001}, $\beta=3.84$ from \citet{Andersen1991}, and with $\beta=3.66\pm0.12$ from \citet{Hohle2010}
The mass function (\autoref{fig:MassFunction_ohneSE}) yields $\Gamma=-2.8\pm0.7$ for stars with $M \gtrsim 8\,M_{\odot}$ and is consistent with $\Gamma=-2.7\pm0.7$ from \citet{Kroupa1993}.
The calculation of mass, age and remaining life-time, like in \autoref{sec:derivedparam}, produces 6~705 stars with a negative remaining life-time. From the subsample of stars with sufficient distances (\autoref{sec:distance}) 2~450 are corrected to positive remaining life-times.
\begin{table}
\centering
\caption{List of $L_{bol}$, mass, age, remaining life-time $\tau_r$, and number N of known additional components of the first ten SN progenitors in \autoref{Tab:Allstars}; with 1\,$\sigma$ errors. Uncertain remaining life-times are marked with a leading colon.}
\begin{tabular}{llrrrrl} \toprule
\# & name              & $L_{bol}$      & mass & age & $\tau_r$ & N \\ 
    &                       & $10^3$ [$L_{\odot}$]  & [$M_{\odot}$] & [Myr] & [Myr] & \\
     \midrule
    1 & $\alpha$ Mus &5.4$\pm$2.4 &8.6(2) &16$\pm$3 &13 &4 \\
    2 & $\eta$ Cen &7.4$\pm$3.5 &8.8(2) &20$\pm$3 &8 &0 \\
    3 & $\beta$ Lup &6.8$\pm$3.2 &8.3(3) &28$\pm$5 &:11 &1 \\
    4 &HR 4898 &3.8$\pm$1.7 &8.9(1) &3$\pm$1 &24 &1 \\
    5 & $\zeta$ Oph &22.4$\pm$2.5 &12.5(1) &14$\pm$1 &2 &0 \\
    6 & $\alpha$ Lup &15.7$\pm$7.5 &9.6(2) &23$\pm$1 &:9 &2 \\
    7 &HR 5193 &6.9$\pm$3.3 &8.7(2) &19$\pm$4 &9 &2 \\
    8 & $\phi$ Cen &4.7$\pm$2.3 &8.4(2) &14$\pm$2 &16 &0 \\
    9 & $\zeta$ Cas &6.3$\pm$3.1 &8.2(2) &28$\pm$5 &7 &0 \\
    10 & $\upsilon$ Sco &16.4$\pm$8.1 &9.7(3) &23$\pm$2 &:4 &0\\
\bottomrule
\end{tabular}
\label{Tab:DerivedParameters_ohneSE}
\end{table}
\newpage

\begin{figure}
\includegraphics[width=\columnwidth, trim=151 143 84 111, clip=true]{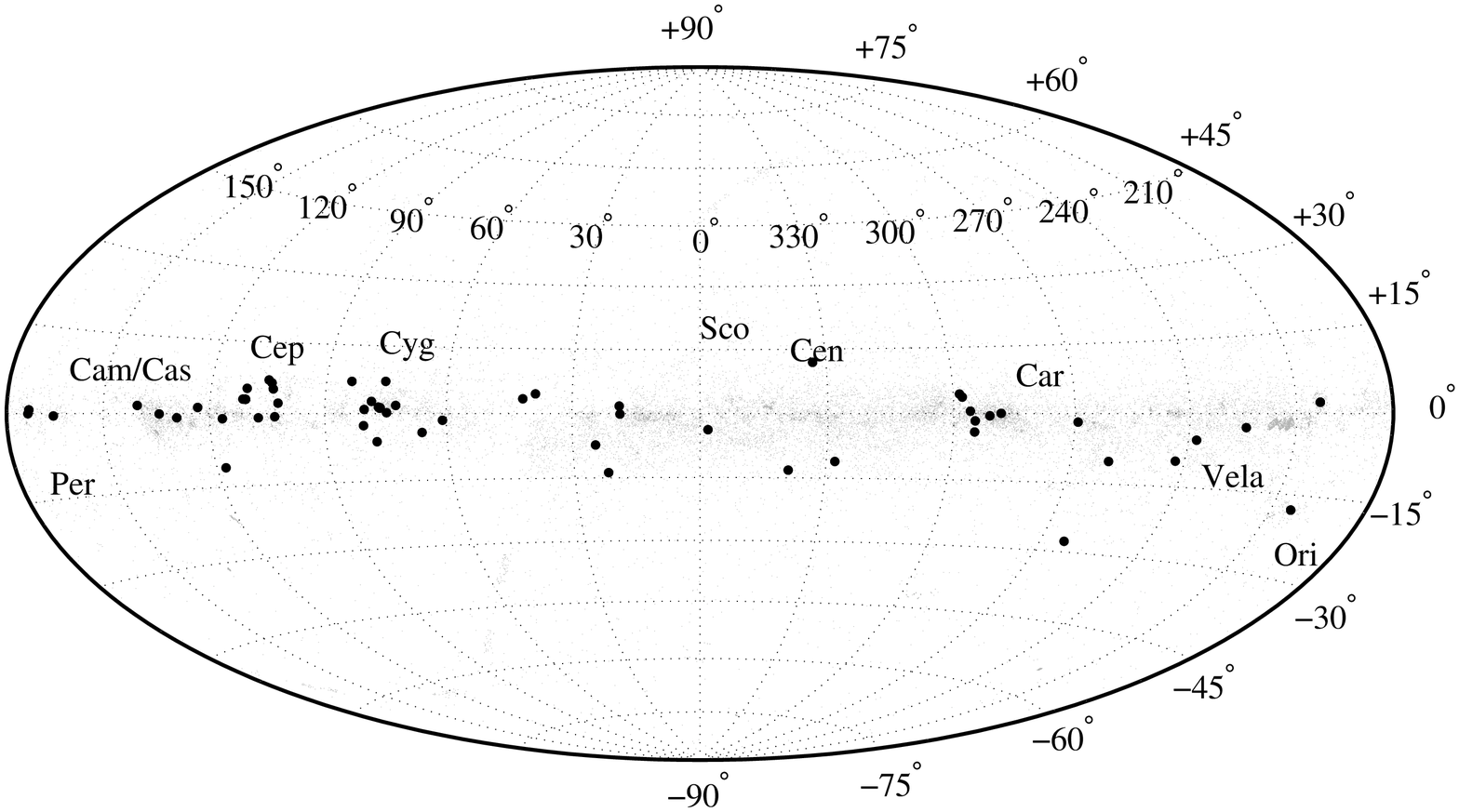}
\caption{The spatial distribution of 130 binaries in galactic coordinates, where both components have at least 8\,$M_{\odot}$ (black dots) compared to the entire sample (grey dots).}
\label{fig:galacticmassivebinaries_ohneSE}
\end{figure}

\begin{figure}
\includegraphics[width=\columnwidth, trim=136 105 86 133,clip=true]{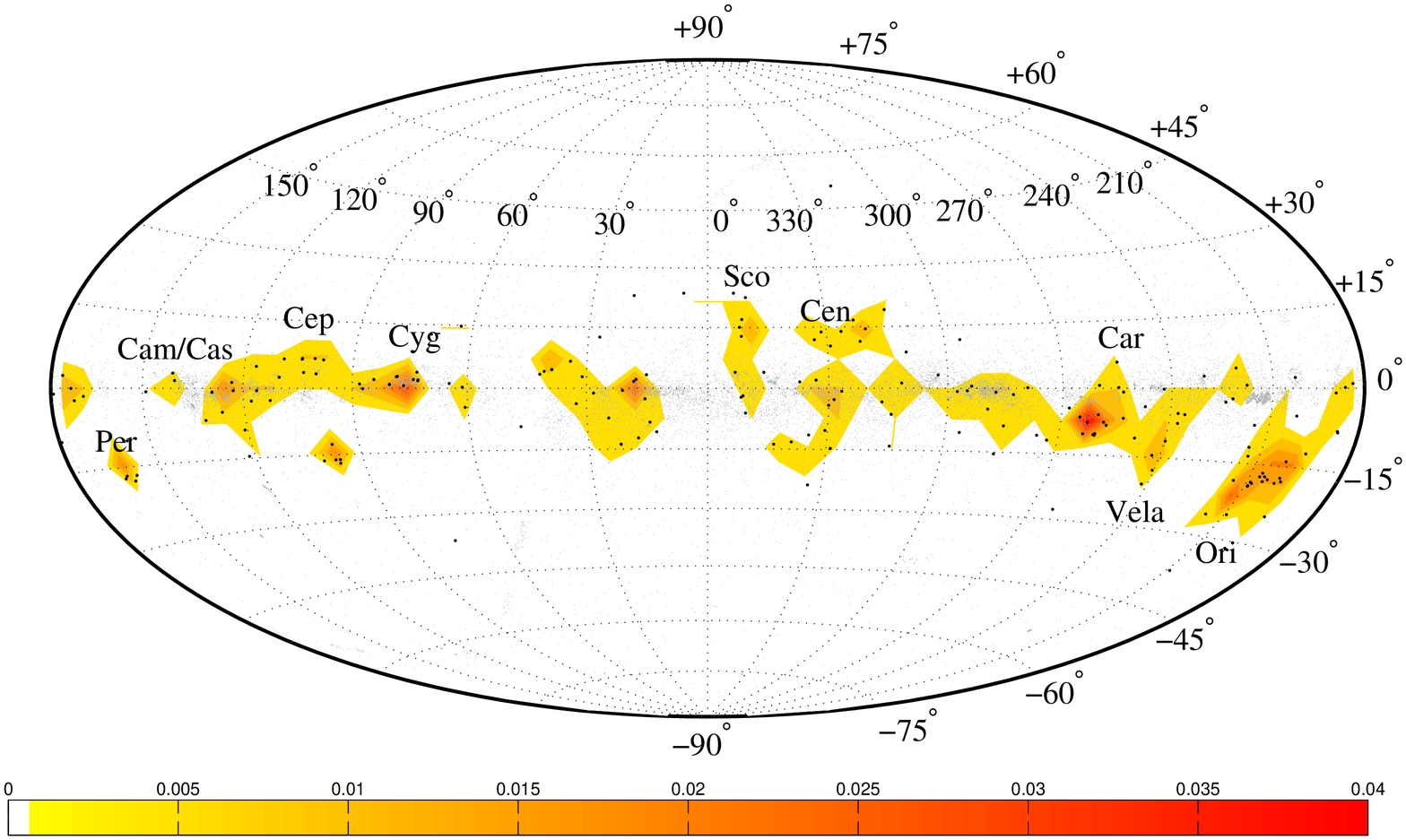}
\caption{The spatial distribution of areas with a local increased SN rate within 600~pc; all SN events will occur in 8.3~\% area of the whole sky and 90~\% SNe in 7.1~\% area of the whole sky. The bin size is 7\,$^\circ\times$7\,$^\circ$.}
\label{fig:snrate600pc_ohneSE}
\end{figure}

\begin{figure}
\includegraphics[width=\columnwidth, trim=133 104 91 126,clip=true]{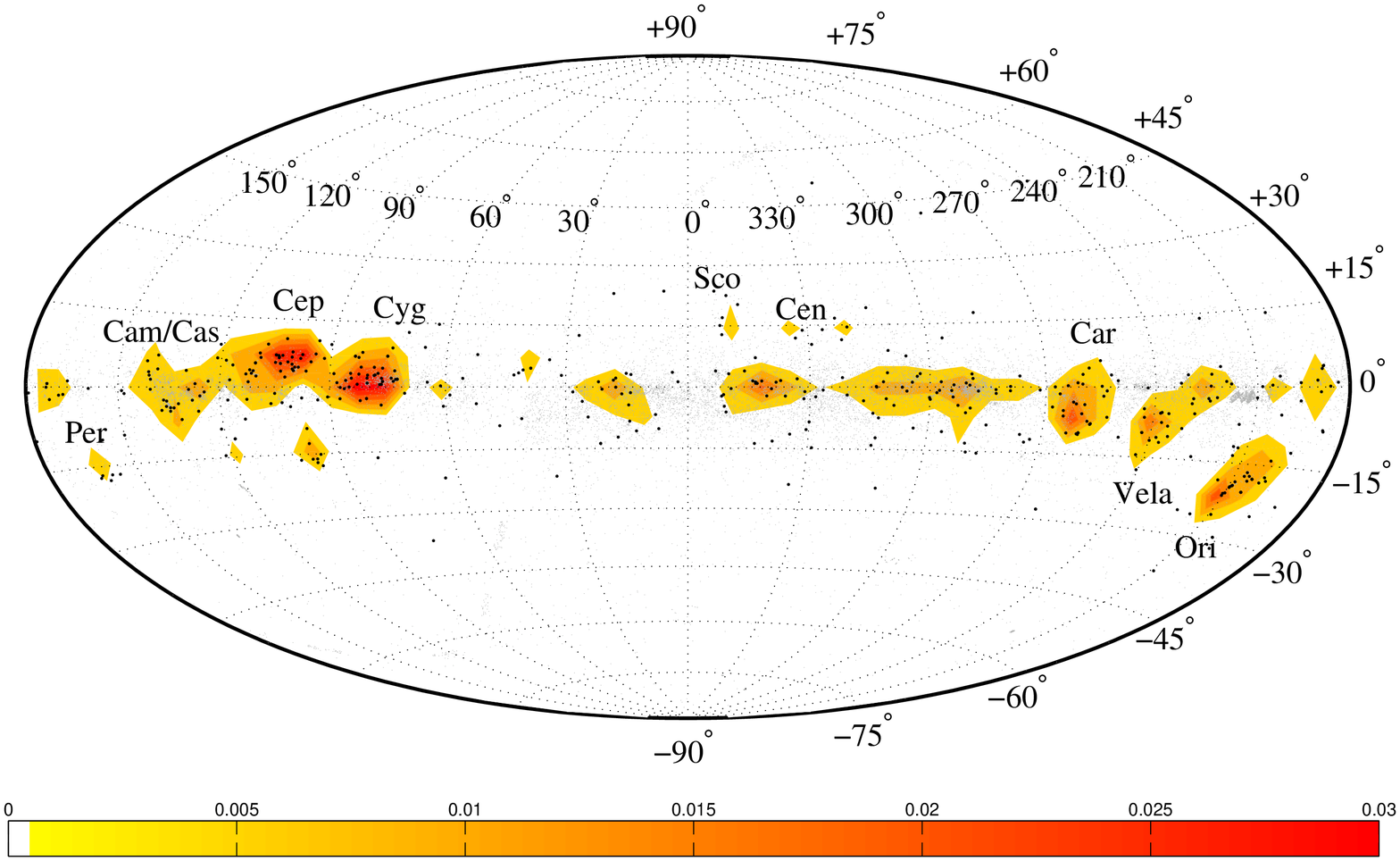}
\caption{The spatial distribution of areas with local increased SN rate within 1~kpc; all SN events will occur in 12.0~\% area of the whole sky and 90~\% SNe in 9.5~\% area of the whole sky. The bin size is 7\,$^\circ\times$7\,$^\circ$.}
\label{fig:snrate1kpc_ohneSE}
\end{figure}

\begin{figure}
\includegraphics[width=\columnwidth, trim=143 107 85 130,clip=true]{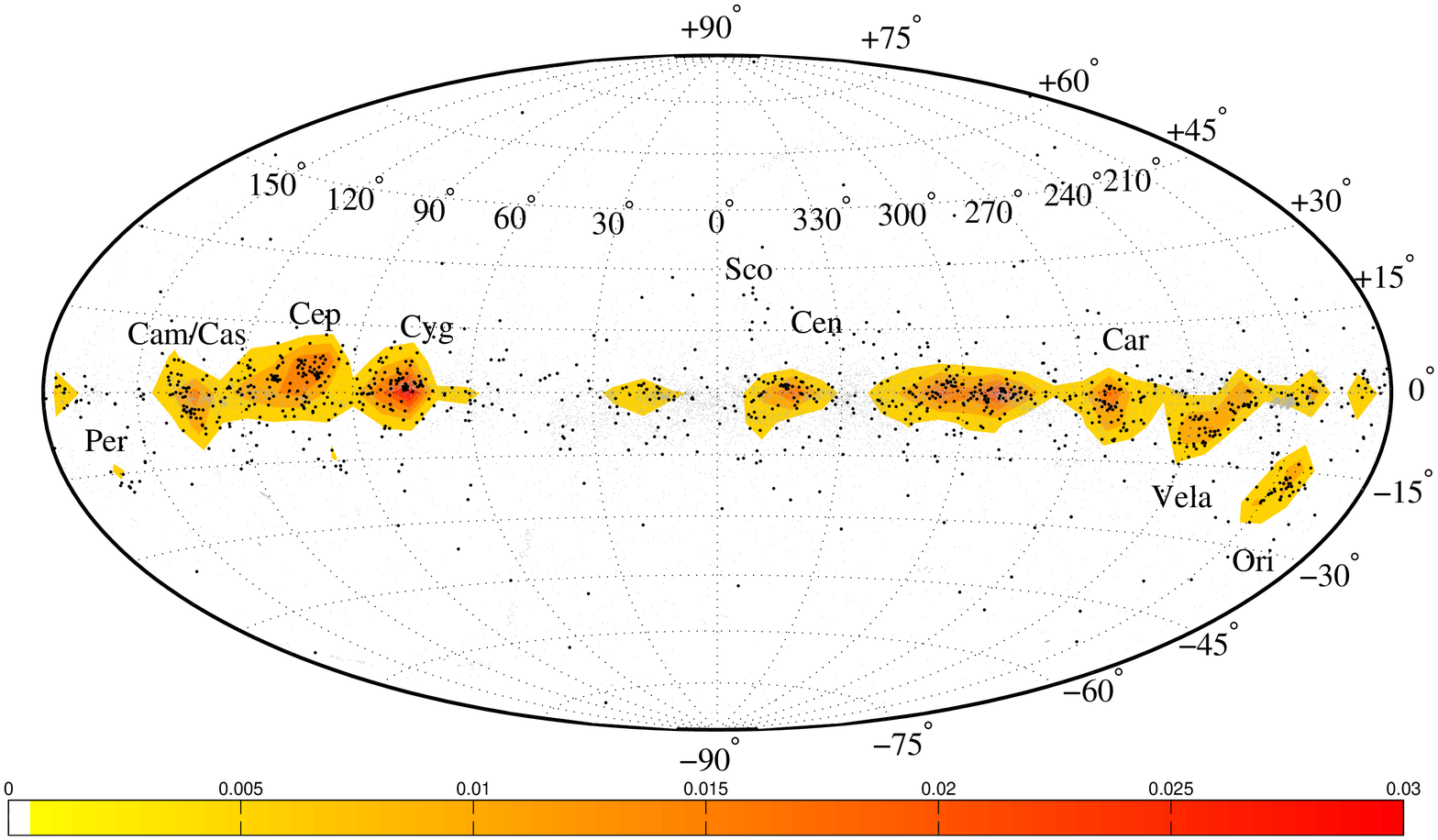}
\caption{The spatial distribution of areas with local increased SN rate within 5~kpc; all SN events in 20.3\% area of the whole sky and 90~\% SNe in 12.1~\% area of the whole sky. The bin size is 7\,$^\circ\times$7\,$^\circ$.}
\label{fig:snrate5kpc_ohneSE}
\end{figure}

\begin{figure}
\includegraphics[width=\columnwidth, trim=155 91 133 115,clip=true]{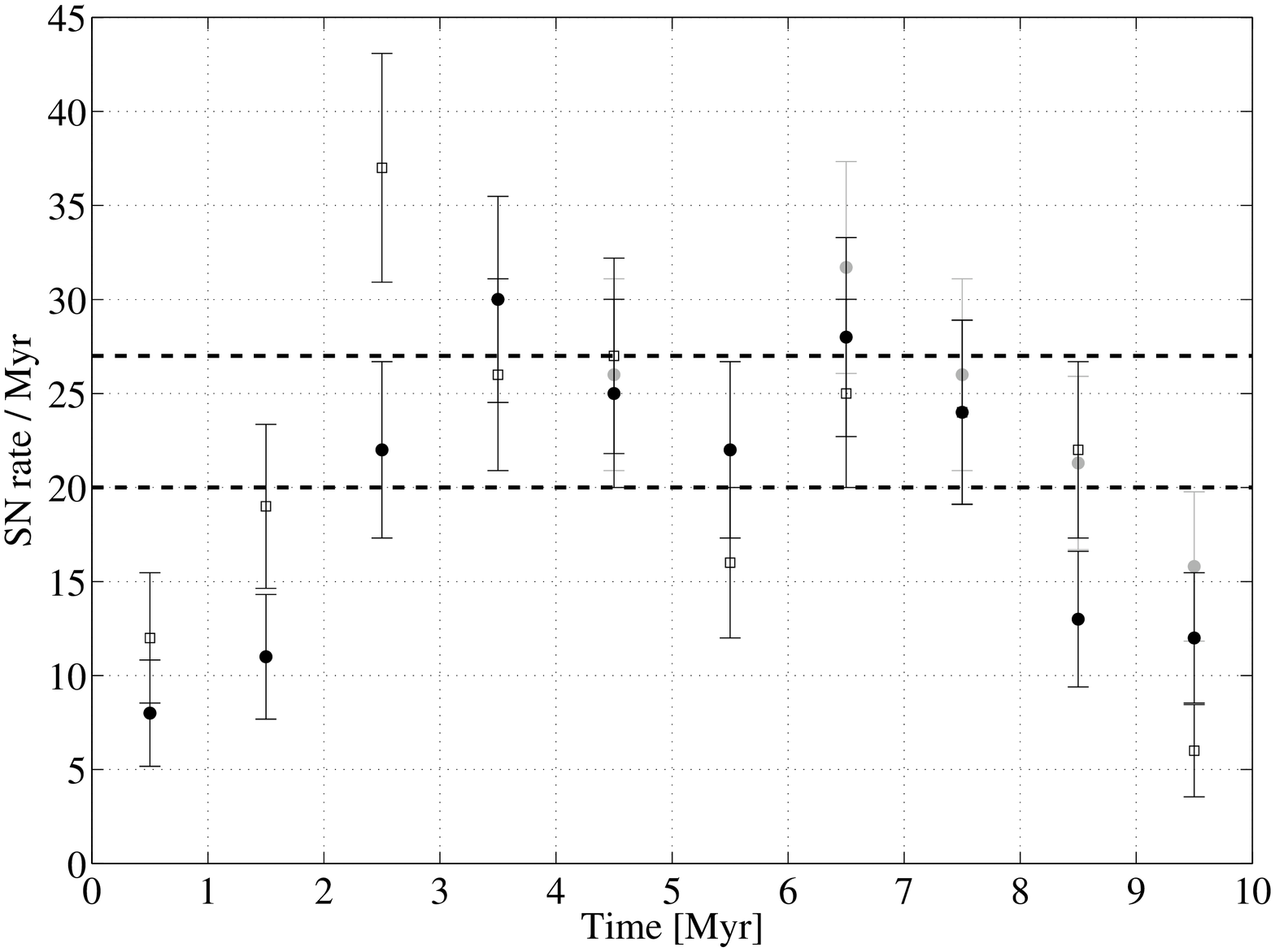}
\caption{The ccSN rate of $19.1^{+4.2}_{-5.4}$ ccSNe/Myr (Poissonian errorbars) from this work for the next 10~Myr within 600~pc (black dots) in comparison with the results in \citet{Hohle2010} (black squares). The dashed line indicates the apparent lower limits for the Gould Belt from \citet{Grenier2004}. The increase of the ccSN rate due to putative massive companions of possible multiple stars is shown by grey dots.}
\label{fig:snratemyr_ohneSE}
\end{figure}

The spatial distribution of 130 binaries, where both components have $M>8\,M_{\odot}$ is plotted in \autoref{fig:galacticmassivebinaries_ohneSE}.
The regions with a locally increased SN rate within a range of up to 600~pc are seen in \autoref{fig:snrate600pc_ohneSE}. All SN events will occur in 8.3~\% area of the whole sky and 90~\% SN events in 7.1~\% area of the whole sky.
The distribution within 1~kpc yields that all SN progenitors are in 12.0~\% of the whole sky and 90~\% SN progenitors are in 9.5~\% of the whole sky (\autoref{fig:snrate1kpc_ohneSE}).
Within 5~kpc (\autoref{fig:snrate5kpc_ohneSE}) all SN progenitors are distributed among 20.3\% area of the whole sky and 90~\% SNe are in 12.1~\% area of the whole sky.
The temporally resolved SN rate of $19.1^{+4.2}_{-5.4}$ (\autoref{fig:snratemyr_ohneSE}) can be compared to the apparent lower limit of 20..27 ccSNe/Myr for the Gould Belt from \citet{Grenier2004} and to $\approx$20 ccSNe/Myr from \citet{Hohle2010}.

\begin{table*}
\centering
\caption{The number of ccSN progenitors with $M>8\,M_{\odot}$ and BH progenitors with $M>30\,M_{\odot}$ within a specific distance. The lower number is obtained by subtracting the 1$\sigma$ error from the masses and the upper value is obtained by adding the 1$\sigma$ error to the masses. The ccSN rate is given for those progenitors which should explode within the next 10\,Myr, the errors are 3$\sigma$. In this table we give the calculated ccSN rate for our Galaxy by extrapolating from the local ccSN rate (given in that row) to the whole Galaxy, with a distance to the Galactic center of 8.5\,kpc, a thick disc scale height of 0.9\,kpc, and a scale length of 3.6\,kpc \citep{Juric2008}. In the last column we give the completeness from \autoref{fig:completeness}.}
\begin{tabular}{llllllll} \toprule
\textbf{distance} & \textbf{SN} & \textbf{BH} & \textbf{SN rate} & \textbf{Gal.SN rate} & \textbf{Completeness}\\ 
		$[$kpc$]$ & [\# progenitors] & [\# progenitors] & [SN/Myr] & [SN/(100~yr)] & [\%]\\ \midrule
      $\leq$\,0.6 & $286^{+35}_{-35}$ & $0^{+0}_{-0}$ & $19.1^{+4.2}_{-5.4}$& $13.4^{+3.0}_{-3.8}$ & 85\\ \hline
	$\leq$\,1.0 & $574^{+51}_{-58}$ & $2^{+1}_{-0}$ & $42.3^{+7.2}_{-9.3}$& $7.3^{+1.2}_{-1.6}$ & 45\\ \hline
	$\leq$\,3.0 & $816^{+84}_{-93}$ & $13^{+4}_{-2}$ & $65.6^{+9.3}_{-17.7}$ & $0.8^{+0.1}_{-0.2}$ & 6\\ \hline		
	$\leq$\,5.0 & $1\,294^{+101}_{-134}$ & $56^{+7}_{-9}$ & $109.9^{+10.5}_{-24.9}$ & $0.4^{+0.1}_{-0.1}$ & 2\\
\bottomrule
\end{tabular}
\label{Tab:SNrate_ohneSE}
\end{table*}

\newpage
\section{Massive binary systems}
\begin{table*}
\centering
\footnotesize
\caption{List of those 35 Binaries$^a$ with at least two massive components, which were also studied in \citet{Hohle2010}, with updated parameters: SIMBAD identifier and distance from \autoref{Tab:Allstars}, spectral types and dynamical mass from [1] \citet{Bondarenko1996}, [2] \citet{Brancewicz1980}, [3] \citet{Docobo2006}, [4] \citet{Perevozkina1999}, [5] \citet{Pourbaix2004}, [6] \citet{Surkova2004}, [7] \citet{Dommanget2002}, and [8] \citet{Mason2010}.}
\tabcolsep=2.7pt
\begin{threeparttable}
\begin{tabular}{cccp{2.6cm}p{2.6cm}cccr} \toprule
  \#  & \textbf{name}   & \textbf{distance} & \multicolumn{2}{c}{\textbf{spectral type}} & \multicolumn{2}{c}{\textbf{mass} $\left[ M_{\odot} \right]$} & \textbf{ref.} \\ 
   & & [pc]         & primary                          & secondary                     & prim.         & sec.     &  \\ \midrule
1& AO Cas & 717$^{+313}_{-167}$ & O7|O9III|O9III	                 &	O8/9|O9III|O9III	        & 20.30-31.73  & 14.80-26.02 & [1][2][5] \\
2&  BM Cas	& 515$^{+360}_{-150}$  &A5I|A5Ia	     	                 &   G0I	                        & 19.88	         & 9.95           & [2][5] \\
3& CC Cas& 553$^{+271}_{-137}$  &	O9.3IV|O9IV	                 & O9IV|O9IV	               	&	20.37	         & 9.98	         &[2][5] \\
4& IU Aur& 216$^{+83}_{-47}$ &	O9.5V|B0:PV| B0|B3|B3Vnne & 	B0.5|B2|B1	            & 12.04-21.30  &	7.94-14.50  & [1][2][5][7][8] \\
5& LY Aur& 252$^{+197}_{-77}$  &	O9.5III|O9.5| O9.5III|B5| O9.5III	& B0IV|O9.5III     &  21.28-24.00 &12.70-18.17	&[1][2][5][7][8]\\
6& $\delta$ Pic& 394$^{+25}_{-22}$  &	B3III|B0.5IV| B0.5III	          & B3.5|B	                &	16.27-16.90 & 8.63-9.00	 & [1][2][5] \\
7& V641 Mon& 586$^{+438}_{-176}$ 	&B1.5IV	                          & B3	                    &	11.50          & 8.40         & [1]  \\
8& FZ CMa& 391$^{+238}_{-107}$  &	B2.5IV-VN| B2.5IV-V           & B2.5IV-V	            & 15.35	         & 15.36	  & [2][5]   \\
9& FF CMa& 827$^{+662}_{-254}$  &	B3	                                  & B4	                    &	11.10	         & 8.88	      & [2]  \\
10& UW CMa&	 579$^{+108}_{-79}$  & O7|O7f|O7f	                      & O7.5|O9III|O7	    & 22.00-32.62  &18.30-24.47 & [1][2][5] \\
11& TU Mus & 1~675$^{+3~367}_{-671}$  &	O7|B3|O7-O8	                  & O9.5|B3               & 22.39-22.60  &15.00-15.40 & [1][2][5] 	\\
12& VZ Cen& 1~555$^{+2~463}_{-591}$  &	B1III:	                              & ?                        &	14.58	         & 	10.21	     & [2]   \\
13& W Cru & 1~922$^{+7~922}_{-857}$  &	G2IAB|G1Iab	                  & ?                        &	11.67	         & 8.18         & [2][5]  \\
14& V701 Sco&  645$^{+517}_{-199}$ &	B1V|B5|B2nn	                  & B1.5                   & 6.42-10.30	 & 5.53-10.2   & [1][2][5]  \\
15& HD168206& 774$^{+710}_{-250}$  &	WC7|WC8                        & B0|O8-9III-V       & 18.49	         & 11.22	   & [2][5]   \\
16& RZ Oph&  437$^{+347}_{-134}$  & F3eIb|F5II-III &	K5II|K5III  &18.46	& 11.14 & [2][5] \\
17& V1182 Aql& 358$^{+164}_{-85}$  &	O9V|O9|O9Vnn                & B1|B3|B3V         & 18.01-38.20	 & 10.85-13.80 & [1][2][5] \\
18& V599 Aql& 214$^{+30}_{-23}$  &	B3.5|B4V|B2V	                 & B3.5|B8              & 18.03-18.40	 & 11.35-11.40 &	[1][2][5] \\
19& $\upsilon$ Sgr& 541$^{+78}_{-60}$  	&B8pI:|B8p         	             & F2pI|F2pe            & 21.26	         & 18.10        & [2][5]  \\
20& V380 Cyg& 643$^{+118}_{-87}$  &	B1III|B1.5II-III	                 & B3V|B2V              & 16.70		     & 9.36	        &	 [2][5] \\
21& V448 Cyg& 650$^{+519}_{-200}$  &	O9.5e|O8.9V| O9.5V	         & B1IB-II|B1.2Ib|B1Ib &  23.84-25.20  & 14.00-15.85 & [2][5][6]  \\
22&V382 Cyg& 674$^{+538}_{-207}$  &O6.5V|O7.5| O6.5|B0|B0    & O7.5|O9|O7.5       & 28.00-37.16	& 19.60-32.69	 &	[1][2][5][7][8] \\
23&V470 Cyg& 589$^{+368}_{-164}$  &B2|B2                             & B2.5|B2              &  13.82	        &12.16            & [2][5]\\
24&V444 Cyg&  604$^{+355}_{-163}$ &O8III:|B1	                     & WN5.5|WN5	        & 34.53	        &19.33	        &	[2][5]\\
25&V729 Cyg& 499$^{+398}_{-153}$  &O7|O7f|O7f| O7e	             & O9-B0|O8|O6f	& 26.70-27.80 & 6.70-23.00	& [1][2][5][8] \\
26&V367 Cyg& 567$^{+181}_{-110}$  &A5EpIA|A5Iab| F5|A7Iape+F5 & A9|A3             & 12.62	        & 8.84	       & [2][5][7][8] \\
27&Y Cyg & 1~118$^{+804}_{-330}$  &B0IV|B0IV	                     & B0IV|B0IV            &17.71	        &	17.54	      & [2][5] \\
28& V1488 Cyg & 321$^{+44}_{-34}$  &  K5IA|K5Iab| K0|K3Ib  & B4V|B4IV-V|B3V                  &	  22.64      &	8.16     & [2][5][7][8] \\
29&HD208392&  485$^{+169}_{-99}$ &B0.5V|B3|B1V	        & B1VE	                 &	10.51	        &	 9.46	    & [2][7][8] \\
30& VV Cep& 744$^{+132}_{-97}$  &M2pIAE| M2epIa|M1| M2Iape& B8VE|B9               & 63.81	        & 35.09	    & [2][5][7][8]   \\
31&HD 211853& 428$^{+221}_{-109}$  &B0:III:|O     	                 & WN6|WN6             & 23.95	        &	16.12	        & [2][5] \\
32&DH Cep& 927$^{+749}_{-286}$  &O5|O6n|B2|O5	             & O5|O6n            & 34.00	        &  27.70	        & [1][5][7][8]  \\
33&AH Cep& 981$^{+519}_{-252}$ & O8|B0.5V| B0.5Vn	             & O9|B0.5|B0.5Vn	 & 15.22-18.10& 13.23-15.90  & [1][2][5]  \\
34&NY Cep& 1~496$^{+1~434}_{-492}$  &B0IV|B0IV| B0|B0IV           & B0IV|B0IV           &  16.07	           & 13.93	    & [2][5][7][8] \\
35&HD218066& 602$^{+358}_{-164}$  &B0.5|B0.5IV-V|B1|B1:V:var& B0.5IV-V| B0.5IV-V& 10.62	           & 9.45	 & [2][5][7][8]  \\
\bottomrule
\end{tabular}
  \begin{tablenotes}
  \item[a] Since there are no coordinates given in \citet{Brancewicz1980} a few \textit{HIPPARCOS} stars were misidentified in Table 5 in \citet{Hohle2010}. We use the correct coordinates in this work.
  \end{tablenotes}
\end{threeparttable}
\label{Tab-A:MassereicheBinaries}
\end{table*}

\begin{table}
\centering
\caption{List of those 6 SN progenitors within 3\,kpc which might be also BH progenitors (see \autoref{Tab:SNrate}): SIMBAD identifier, distance, and spectral type from \autoref{Tab:Allstars}, mass and remaining life-time $\tau_r$ from \autoref{Tab:DerivedParameters}.}
\begin{tabular}{lllllll}\toprule
\textbf{name}	&	\textbf{spec.type}	&\textbf{distance}	& \textbf{mass}	& {\boldmath$\tau_r$}\\
 						&  				&[kpc]									& [$M_\odot$]	& [Myr] \\ \midrule
V973 Sco			& O8Iaf		&$1.7^{+1.1}_{-0.5}$		& 35.0$\pm$1.7	& 2.5 \\
V453 Sco			& B1Ia(e)p	&$1.4^{+1.2}_{-0.5}$		& 31.0$\pm$0.3	& 2.3 \\
9 Sgr				& O4V			&$1.6^{+1.3}_{-0.5}$		& 73$\pm$14		& 2.8 \\
HD 14434		& O5.5V		&$2.8^{+13}_{-1.4}$		& 33.5$\pm$8.5	& 4.0\\
Naos				& O6/7III		&$0.33\pm{0.01}$				& 35.6$\pm$3.8	& 2.0\\
VV Cep			&M3:Ia-Iab:	&$0.7\pm{0.1}$					& 63.8$\pm$0.6	& 2.4\\
\bottomrule
\end{tabular}
\label{Tab-A:BHprogenitors}
\end{table}
\label{lastpage}

\begin{thebibliography}{99}
\bibitem[\protect\citeauthoryear{Aasi et al.}{2013}]{Aasi2013} Aasi, J., Abadie, J., Abbott, B.~P., et al.: 2013, PhRvD, 87, 042001 
\bibitem[\protect\citeauthoryear{Ackermann et al.}{2013}]{Ackermann2013} Ackermann, M., Ajello, M., Allafort, A., et al.: 2013, Science, 339, 807 
\bibitem[\protect\citeauthoryear{Andersen}{1991}]{Andersen1991}{Andersen}, J.: 1991, A\&ARv 3, 91
\bibitem[\protect\citeauthoryear{Bertelli et al.}{1994}]{Bertelli1994}{Bertelli}, G., {Bressan}, A., {Chiosi}, C., et al.: 1994, A\&AS 106, 275
\bibitem[\protect\citeauthoryear{Bessell et al.}{1998}]{Bessell1998}{Bessell}, M.~S., {Castelli}, F., {Plez}, B.: 1998, A\&A 333, 231
\bibitem[\protect\citeauthoryear{Bessell et al.}{1988}]{Bessell1988}{Bessell}, M.~S., {Brett}, J.~M.: 1988, PASP 100, 1134
\bibitem[\protect\citeauthoryear{Bondarenko \& Perevozkina}{1996}]{Bondarenko1996}{Bondarenko}, I.~I., {Perevozkina}, E.~L.: 1996, Bondarenko, I.~I., \& Perevozkina, E.~L.: 1996, Odessa Astronomical Publications 9, 20 
\bibitem[\protect\citeauthoryear{Brancewicz \& Dworak}{1980}]{Brancewicz1980}{Brancewicz}, H.~K., {Dworak}, T.~Z.: 1980, ACTAA 30, 501
\bibitem[\protect\citeauthoryear{Cardelli et al.}{1989}]{Cardelli1989}{Cardelli}, J.~A., {Clayton}, G.~C., {Mathis}, J.~S.: 1989, ApJ 345, 245
\bibitem[\protect\citeauthoryear{Claret}{2004}]{Claret2004}{Claret}, A.: 2004, A\&A~424, 919
\bibitem[\protect\citeauthoryear{de Bruijne}{2012}]{deBruijne2012} de Bruijne, J.~H.~J.: 2012, ApSS 341, 31 
\bibitem[\protect\citeauthoryear{Diehl et al.}{2006}]{Diehl2006} {Diehl}, R., {Halloin}, H., 
{Kretschmer}, K., et al.: 2006, Nat. 439, 45 
\bibitem[\protect\citeauthoryear{Docobo \& Andrade}{2006}]{Docobo2006}{Docobo}, J.~A., {Andrade}, M.: 2006, ApJ 652, 681
\bibitem[\protect\citeauthoryear{Docobo \& Andrade}{2013}]{Docobo2013} Docobo, J.~A., \& Andrade, M.: 2013, MNRAS 428, 321 
\bibitem[\protect\citeauthoryear{Dommanget \& Nys}{2002}]{Dommanget2002}{Dommanget}, J., {Nys}, O.: 2002, CCDM (Catalog of Components of Double \& Multiple stars), VizieR Online Data Catalog 1274
\bibitem[\protect\citeauthoryear{Gould}{1879}]{Gould1879} {Gould}, B.~G.: 1879, Resultados del Observatorio Nacional Argentino, 1, 14
\bibitem[\protect\citeauthoryear{Grenier}{2004}]{Grenier2004}{Grenier}, I.~A.: 2004,
arXiv, astro-ph/0409096
\bibitem[\protect\citeauthoryear{Harding et al.}{2001}]{Harding2001} Harding, P., Morrison, H.~L., Olszewski, E.~W., et al.: 2001: AJ, 122, 1397 
\bibitem[\protect\citeauthoryear{Hilditch}{2001}]{Hilditch2001}{Hilditch}, R.~W.: 2001,
An Introduction to Close Binary Stars
\bibitem[\protect\citeauthoryear{Hoeg et al.}{1997}]{Hoeg1997}{Hoeg}, E., {B{\"a}ssgen}, G., {Bastian}, U., et al.: 1997, A\&A 323, L49
\bibitem[\protect\citeauthoryear{Hohle, Neuh\"auser \& Schutz}{Hohle et al.}{2010}]{Hohle2010}{Hohle}, M.~M., {Neuh\"auser}, R., {Schutz}, B.F.: 2010, AN 331, 349
\bibitem[\protect\citeauthoryear{Juri{\'c} et al.}{2008}]{Juric2008} Juri{\'c}, M., 
Ivezi{\'c}, {\v Z}., Brooks, A., et al.: 2008, ApJ 673, 864 
\bibitem[\protect\citeauthoryear{Kenyon \& Hartmann}{1995}]{Kenyon1995}{Kenyon}, S.~J., {Hartmann}, L.: 1995, ApJS 101, 117
\bibitem[\protect\citeauthoryear{Kharchenko \& Roeser}{2009}]{Kharchenko2009}{Kharchenko}, N.~V., {Roeser}, S.: 2009, All-sky Compiled Catalogue of 2.5 million stars, VizieR Online Data Catalog 1280
\bibitem[\protect\citeauthoryear{Kodama}{1997}]{Kodama1997}{Kodama}, T.: 1997, PhD thesis, Institute of Astronomy, Univ. Tokyo
\bibitem[\protect\citeauthoryear{Kroupa et al.}{1993}]{Kroupa1993}{Kroupa}, P., {Tout}, C.~A., {Gilmore}, G.: 1993, MNRAS 262, 545
\bibitem[\protect\citeauthoryear{Lang}{1992}]{Lang1992}{Lang}, K.~R.: 1992, {Astrophysical Data I. Planets and Stars.}
\bibitem[\protect\citeauthoryear{Lynga}{1982}]{Lynga1982}{Lynga}, G.: 1982, A\&A 109, 213 
\bibitem[\protect\citeauthoryear{Maeder \& Meynet}{1989}]{Maeder1989}{Maeder}, A., {Meynet}, G.: 1989, A\&A 210, 155
\bibitem[\protect\citeauthoryear{Manchester et al.}{2005}]{Manchester2005} Manchester, R.~N., Hobbs, G.~B., Teoh, A., \& Hobbs, M.: 2005, AJ 129, 1993 
\bibitem[\protect\citeauthoryear{Mason et al.}{2010}]{Mason2010}{Mason}, B.~D., {Wycoff}, G.~L., {Hartkopf}, W.~I., et al.: 2010, The Washington Visual Double Star Catalog, VizieR Online Data Catalog 1020
\bibitem[\protect\citeauthoryear{Meynet \& Maeder}{2003}]{Meynet2003}{Meynet}, G. and {Maeder}, A.: 2003, A\&A 404, 975
\bibitem[\protect\citeauthoryear{Mihalas \& Binney}{1981}]{Mihalas1981} Mihalas, D., \& Binney, J.: 1981: San Francisco, CA, W.~H.~Freeman and Co., 1981.~608 p.,  
\bibitem[\protect\citeauthoryear{Ochsenbein et al.}{2000}]{Ochsenbein2000}{Ochsenbein}, F., {Bauer}, P., {Marcout}, J.: 2000, A\&AS 143, 23
\bibitem[\protect\citeauthoryear{Oudmaijer \& Parr}{2010}]{Oudmaijer2010} Oudmaijer, R.~D., \& Parr, A.~M.: 2010, MNRAS 405, 2439 
\bibitem[\protect\citeauthoryear{Palomba}{2005}]{Palomba2005} {Palomba}, C.: 2005, MNRAS 359, 1150 
\bibitem[\protect\citeauthoryear{Perevozkina \& Svechnikov}{1999}]{Perevozkina1999}{Perevozkina}, E.~L., {Svechnikov}, M.~A.: 1999, Catalog of eclipsing binaries parameters, {VizieR Online Data Catalog} 5118
\bibitem[\protect\citeauthoryear{Perryman et al.}{1997}]{Perryman1997}{Perryman}, M.~A.~C., {Lindegren}, L., {Kovalevsky}, J., et al.: 1997, A\&A 323, L49 
\bibitem[\protect\citeauthoryear{Pires et al.}{2009}]{Pires2009}{Pires}, A.~M. and {Motch}, C. and {Turolla}, R. and {Treves}, A. and {Popov}, S.~B.,: 2009 A\&A 498, 233
\bibitem[\protect\citeauthoryear{Popov et al.}{2005}]{Popov2005}Popov, S.~B., Turolla, R., Prokhorov, M.~E., Colpi, M., \& Treves, A.: 2005, ApSS 299, 117 
\bibitem[\protect\citeauthoryear{Pourbaix et al.}{2004}]{Pourbaix2004}{Pourbaix}, D., {Tokovinin}, A.~A., {Batten}, A.~H., et al.: 2004, A\&A 424, 727
\bibitem[\protect\citeauthoryear{Reed}{2005}]{Reed2005} Reed, B.~C.: 2005, AJ, 130, 1652 
\bibitem[\protect\citeauthoryear{Rieke \& Lebofsky}{1985}]{Rieke1985}{Rieke}, G.~H., {Lebofsky}, M.~J.: 1985, ApJ 288, 618
\bibitem[\protect\citeauthoryear{Savage \& Mathis}{1979}]{Savage1979}{Savage}, B.~D., {Mathis}, J.~S.: 1979, ARAA 17, 73
\bibitem[\protect\citeauthoryear{Schaller et al.}{1992}]{Schaller1992}{Schaller}, G., {Schaerer}, D., {Meynet}, G., et al.: 1992, A\&AS 96, 269
\bibitem[\protect\citeauthoryear{Schroeder et al.}{2004}]{Schroeder2004}{Schr{\"o}der}, S.~E., {Kaper}, L., {Lamers}, H.~J.~G.~L.~M., et al.: 2004, A\&A 428, 149
\bibitem[\protect\citeauthoryear{Smith \& Eichhorn}{1996}]{Smith1996}{Smith}, Jr. H., {Eichhorn}, H.: 1996, MNRAS 281, 211
\bibitem[\protect\citeauthoryear{Skiff}{2013}]{Skiff2013}{Skiff}, B.~A.: 2013, Catalogue of Stellar Spectral Classifications, {VizieR Online Data Catalog} 2023
\bibitem[\protect\citeauthoryear{Skrutskie et al.}{2006}]{Skrutskie2006}Skrutskie, M.~F., 
Cutri, R.~M., Stiening, R., et al.: 2006, AJ~131, 1163
\bibitem[\protect\citeauthoryear{Stothers \& Frogel}{1974}]{Stothers1974} Stothers, R., \& Frogel, J.~A.: 1974, AJ 79, 456 
\bibitem[\protect\citeauthoryear{Sota et al.}{2011}]{Sota2011}{Sota}, A., {Ma{\'{\i}}z Apell{\'a}niz}, J., {Walborn}, N.~R. et al.: 2011, ApJS~193,24
\bibitem[\protect\citeauthoryear{Surkova \& Svechnikov}{2004}]{Surkova2004}{Surkova}, L.~P., {Svechnikov}, M.~A.: 2004, Semi-detached eclipsing binaries, {VizieR Online Data Catalog} 5115
\bibitem[\protect\citeauthoryear{Svensmark et al.}{2012}]{Svensmark2012}{Svensmark}, J., {Enghoff}, M.~B., {Svensmark}, H.: 2012, ACPD 12, 3595
\bibitem[\protect\citeauthoryear{Tammann et al.}{1994}]{Tammann1994}{Tammann}, G.~A., {Loeffler}, W., {Schroeder}, A.: 1994, ApJS 92, 487
\bibitem[\protect\citeauthoryear{Torres}{2010}]{Torres2010} Torres, G.: 2010, AJ 140, 1158 
\bibitem[\protect\citeauthoryear{van der Hucht}{2001}]{vanderHucht2001}{{van der Hucht}, K.~A.}: 2001, NewAR 45, 135
\bibitem[\protect\citeauthoryear{van Leeuwen}{2007}]{vanLeeuwen2007}{van Leeuwen}, F.: 2007, A\&A 474, 653
\bibitem[\protect\citeauthoryear{Wenger et al.}{2000}]{Wenger2000}{Wenger}, M., {Ochsenbein}, F., {Egret}, et al.: 2000, A\&AS 143, 9
\bibitem[\protect\citeauthoryear{Wegner et al.}{2008}]{Wegner2007}{Wegner}, W.: 2007, MNRAS~374, 1549
\bibitem[\protect\citeauthoryear{Woosley et al.}{2002}]{Woosley2002} Woosley, S.~E., Heger, A., \& Weaver, T.~A.: 2002, Reviews of Modern Physics, 74, 1015 
\bibitem[\protect\citeauthoryear{Zacharias et al.}{2005}]{Zacharias2005}{Zacharias}, N., {Monet}, D.~G., {Levine}, S.~E., et al.: 2005, NOMAD Catalog, {VizieR Online Data Catalog} 1297
\bibitem[\protect\citeauthoryear{Zacharias et al.}{2013}]{Zacharias2013}{Zacharias}, N., {Finch}, C.~T., {Girard}, T.~M., et al.: 2013, AJ 145, 44 
\bibitem[\protect\citeauthoryear{Zinnecker \& Yorke}{2007}]{Zinnecker2007}{Zinnecker}, H. and {Yorke}, H.~W.: 2007, ARAA 45, 481
\end{thebibliography}
\end{document}